\documentclass[amsmath,amssymb,twocolumn,showpacs,superscriptaddress,groupedaddress,aps,prd,10pt,nofootinbib]{revtex4-1}
\usepackage{ae}
\usepackage{bm} 
\usepackage{color}
\usepackage{amssymb}
\usepackage{amsmath}
\usepackage{amsfonts}
\usepackage{mathtools}
\usepackage{graphicx}
\usepackage{slashed}
\usepackage{wasysym}
\usepackage{setspace}
\usepackage{subfigure}
\usepackage{ulem}

\usepackage{tabularx}
\usepackage{graphicx}
\usepackage[boxed]{algorithm}
\usepackage{algpseudocode}
\usepackage{nicefrac}
\usepackage{booktabs}
\usepackage{tikz}
\usepackage{placeins}

\makeatletter
\@ifclasswith{revtex4-1}{preprint}
{
\newcount\Bool	

\allowdisplaybreaks

\linespread{1.308}
\newlength{\figlenD}
\newlength{\figlenN} 
\newlength{\figlenPerp}
\newlength{\figlenFull}
\newlength{\figlenCut}
\newlength{\figlendN}

\figlenD=0.40\textwidth
\figlenN=0.42\textwidth
\figlenPerp=0.525\textwidth
\figlenFull=0.50\textwidth
\figlenCut=0.49\textwidth
\figlendN=0.6\textwidth

\def\eq$#1${\begin{equation}#1\end{equation}} 
\newcommand{\br}{}

}
{
\newcount\Bool

\newlength{\figlenD}
\newlength{\figlenN} 
\newlength{\figlenSph} 
\newlength{\figlenSphh} 
\newlength{\figlenPerp}
\newlength{\figlenFull}
\newlength{\figlenCut}
\newlength{\figlendN}

\figlenD=0.42\textwidth
\figlenN=0.38\textwidth
\figlenSph=0.32\textwidth
\figlenSphh=0.36\textwidth
\figlenPerp=0.525\textwidth
\figlenFull=0.48\textwidth
\figlenCut=0.49\textwidth
\figlendN=0.58\textwidth

\def\eq$#1${\begin{multline}#1\end{multline}}
\newcommand{\br}{\\}

\widowpenalty10000
\clubpenalty10000
\brokenpenalty500\relax
\hyphenpenalty10000
\interfootnotelinepenalty=10000
}
\makeatother

\newcommand{\Eqref}[1]{Eq.~\eqref{#1}}

\newcommand*{\I}{ {\rm i} }

\floatname{algorithm}{Procedure}

\begin{document}

\setlength{\unitlength}{1mm}
\title{All-optical signatures of Strong-Field QED in the vacuum emission picture\vspace*{-1.3mm}}
\author{Holger Gies}\email{holger.gies@uni-jena.de}
\author{Felix Karbstein}\email{felix.karbstein@uni-jena.de}
\author{Christian Kohlf\"urst}\email{christian.kohlfuerst@uni-jena.de}
\affiliation{Helmholtz-Institut Jena, Fr\"obelstieg 3, 07743 Jena, Germany}
\affiliation{Theoretisch-Physikalisches Institut, Abbe Center of Photonics, \\ 
Friedrich-Schiller-Universit\"at Jena, Max-Wien-Platz 1, 07743 Jena, Germany}

\date{\today}

\begin{abstract}
  We study all-optical signatures of the effective nonlinear couplings
  among electromagnetic fields in the quantum vacuum, using the
  collision of two focused high-intensity laser pulses as an example.
  The experimental signatures of quantum vacuum nonlinearities are
  encoded in signal photons, whose kinematic and polarization
  properties differ from the photons constituting the macroscopic
  laser fields.  We implement an efficient numerical algorithm
  allowing for the theoretical investigation of such signatures in
  realistic field configurations accessible in experiment.  This
  algorithm is based on a vacuum emission scheme and can readily be
  adapted to the collision of more laser beams or further involved
  field configurations. We solve the case of two colliding pulses in full 3+1 dimensional spacetime, and
  identify experimental geometries and parameter regimes with improved
  signal-to-noise ratios.
\end{abstract}

\maketitle

\section{Introduction}

The fluctuations of virtual particles in the quantum vacuum gives rise
to effective interactions among electromagnetic fields, supplementing
Maxwell's linear theory of vacuum electrodynamics with effective
nonlinearities \cite{Heisenberg:1935qt,Weisskopf,Schwinger:1951nm};
for reviews, see
Refs.~\cite{Dittrich:1985yb,Dittrich:2000zu,Marklund:2008gj,Dunne:2008kc,Heinzl:2008an,DiPiazza:2011tq,Dunne:2012vv,Battesti:2012hf,King:2015tba,Karbstein:2016hlj}.
Prominent signatures of quantum vacuum nonlinearities are vacuum
magnetic birefringence (VMB) \cite{Toll:1952,Baier} and direct
light-by-light scattering \cite{Euler:1935zz,Karplus:1950zza}.

Being of quantum nature, the latter are typically tiny and rather
elusive in experiment.  In quantum electrodynamics (QED), they are
suppressed parametrically with inverse powers of the electron mass
$m_e$.  This mass scale serves as the typical energy to be compared
with the scales of the applied fields, and defines the {\it critical
  field strengths} $E_\text{cr}:=\frac{c^3}{\hbar}\frac{m_e^2}{e}
\approx 1.3 \times 10^{16} \frac{\rm V}{\rm cm}$ and
$B_\text{cr}:=\frac{E_\text{cr}}{c} \approx 4 \times 10^{9} {\rm T}$.

In the laboratory, field strengths of this order are only reached in
strong Coulomb fields of highly charged ions. Hence,
experimental verifications of QED vacuum nonlinearities have so far
been limited to high-energy experiments with highly charged ions
\cite{Lee:1998hu,Akhmadaliev:1998zz,Akhmadaliev:2001ik,dEnterria:2013zqi,Aaboud:2017bwk}.
Note, that VMB is potentially also relevant for the optical polarimetry of neutron stars \cite{Mignani:2016fwz,Capparelli:2017mlv,Turolla:2017tqt}.
Even though QED vacuum nonlinearities in macroscopic electromagnetic fields have
not been directly verified yet, laboratory searches of VMB in macroscopic magnetic fields
\cite{Cantatore:2008zz,Berceau:2011zz,Fan:2017fnd} have already
demonstrated the need for high field strengths and, at the same time,
a high signal detection sensitivity, see also
\cite{Zavattini:2016sqz,Inada:2017}.  The demand for strong
fields together with the recent technological advances in the
development of high-intensity laser systems have opened up an
alternative route to access the extreme-field territory in the
laboratory. The overarching key idea is to combine high-intensity
lasers with polarization sensitive single photon detection schemes.

State-of-the-art high-intensity lasers reach peak field strengths of
the order of $10^6{\rm T}$ and $10^{12}{\rm V}/{\rm cm}$ in micron
sized focal spots.  Laser pulses achieving these field
strengths are typically made up of ${\cal O}(10^{20})$ photons,
constituting a challenging background for the detection of the
generically tiny signals of QED vacuum nonlinearities in experiment.
In this context, theoretical proposals specifically focused on
VMB \cite{Heinzl:2006xc,DiPiazza:2006pr,Dinu:2013gaa,Karbstein:2015xra,Schlenvoigt:2016,Karbstein:2016lby,King:2016jnl,Bragin:2017yau},
photon-photon scattering in the form of laser-pulse collisions
\cite{Lundstrom:2005za,Lundin:2006wu,King:2012aw}, quantum reflection
\cite{Gies:2013yxa,Gies:2014wsa}, photon merging
\cite{Yakovlev:1966,DiPiazza:2007cu,Gies:2014jia,Gies:2016czm} and
splitting
\cite{Adler:1970gg,BialynickaBirula:1970vy,Adler:1971wn,Papanyan:1971cv,DiPiazza:2007yx},
and optical signatures of QED vacuum nonlinearities based on
interference effects
\cite{King:2013am,Tommasini:2010fb,Hatsagortsyan:2011}.

In this article, we introduce and benchmark an efficient numerical
algorithm tailored to the study of all-optical signatures of QED
vacuum nonlinearities.  Reformulating the signatures in terms of {\it
  vacuum emission} processes \cite{Karbstein:2014fva}, the effects of
quantum vacuum nonlinearities are encoded in signal photons emitted
from the strong-field region. As no signal photons are induced in the
absence of vacuum nonlinearities, these photons generically constitute
a distinct signal.  However, in order to allow for their detection in
experiment, they have to differ from the photons constituting the
high-intensity laser pulses driving the effects, e.g., by their
kinematic and polarization properties.  Correspondingly, one central
objective is to identify scenarios where such effects are most
pronounced.

A standard approach of dealing with this challenge is to solve the
nonlinear photon wave equation, i.e. a partial differential equation,
by suitable numerical techniques. Successful examples can be found,
e.g., in \cite{King:2012aw}, where the nonlinearities of the
field equations have been treated as source terms and Green's function
methods are used for an iterative solution strategy; see also
\cite{King:2014vha,Bohl:2015uba} for an advanced implementation based
on the pseudo-characteristic method of lines. For large-scale
simulation purposes, an implicit ODE-based solver has been
specifically designed in \cite{Domenech:2016xx}, as well as in
\cite{Carneiro:2016qus} using a finite-difference time-domain solver.

As demonstrated in the following, the vacuum emission picture
advocated in this work is particularly suited for a numerical
implementation.  In our formalism, the essential numerical ingredients
are reduced to one standard and easy-to-use algorithm: fast Fourier
transformation. Space- or time-integrated observables may additionally
require simple low-dimensional integration techniques. This numerical
simplicity parallels the conceptual adaption of the vacuum emission
scheme to the physical situation: in this picture, all macroscopically
controlled fields such as high-intensity laser pulses are treated as
classical fields, whereas the fluctuation-induced signal photons are
dealt with on the level of the quantum Fock space. 

Our article is organized as follows: In Sec.~\ref{sec:theofond} we
outline the theoretical foundations of our approach. We apply our
methods in Sec.~\ref{sec:2hilpcollision} to the collision of two
focused, linearly polarized high-intensity laser pulses in vacuum
\cite{King:2012aw}. In Sec.~\ref{sec:numerics}, we introduce our
numerical algorithm in detail. Section~\ref{sec:results} is devoted to
the discussion of explicit results. Here, we first benchmark our
numerical algorithm with analytical results for the limit of infinite
Rayleigh ranges of the two beams, where analytical results are
available.  Subsequently, we use it to obtain new results: in
Sec.~\ref{subsec:collA}, we study the collision of two petawatt class
laser pulses of identical frequency, continuing with fundamental and
doubled frequency in Sec.~\ref{subsec:collB}.  Considering the
fundamental frequency laser beam as focused down to the diffraction
limit, the latter scenario allows for the study of two limiting cases
of specific interest, differing in the focusing of the
frequency-doubled beam. In the first case, it is focused to the
diffraction limit of the fundamental frequency beam, maximizing the
beam overlap in the focus, and in the second case to its own
diffraction limit, resulting in a narrower beam waist and thus in a
considerably smaller overlap region of the beams but higher intensity
in the focus.  Finally, we end with conclusions and an outlook in
Sec.~\ref{sec:concl}.

\section{Theoretical foundations}\label{sec:theofond}

In Ref.~\cite{Karbstein:2014fva}, it has been argued that all-optical 
signatures of quantum vacuum nonlinearities can be efficiently analyzed 
by reformulating them in terms of {\it vacuum emission} processes.
This approach has meanwhile been successfully employed to obtain experimentally 
realistic predictions for the phenomenon of VMB, particularly 
in the combination of x-ray free electron and high-intensity lasers 
\cite{Karbstein:2015xra,Karbstein:2015qwa,Karbstein:2016lby}. 

The central idea is to consider all applied macroscopic
electromagnetic fields as constituting the external background field;
cf also Ref. \cite{Gies:2016yaa}.  This implies, that the quantum
character of the applied fields is not resolved, and effects like,
e.g., QED-induced beam depletion are neglected.  We
emphasize that this is typically well-justified for scenarios where
the strong electromagnetic fields $E$ and $B$ are provided by
high-intensity lasers and fulfill $E\ll E_\text{cr}$ and $B\ll
B_\text{cr}$.  Due to the parametric suppression of QED vacuum
nonlinearities by powers of the electron mass, the pulses delivered by
such lasers can be considered as traversing each other in vacuum
essentially unaltered.

At one-loop order, but fully nonperturbative in the background field $\bar A$, 
the {\it exact} interaction term giving rise to single signal photon emission is 
given by \cite{Gies:2016yaa}
\begin{equation}
 \Gamma_\text{int}^{(1)}[\bar A(x)]=\int{\rm 
d}^4x\,\frac{\delta\Gamma^{1\text{-loop}}_\text{HE}[A]}{\delta 
A^\mu}\bigg|_{A=\bar A(x)}a^\mu(x) \,,
\end{equation}
where $\Gamma^{1\text{-loop}}_\text{HE}[A]=-\I \ln\det(-\I
\slashed{\partial}-e\slashed{A} +m_e)$ is the one-loop
Heisenberg-Euler action evaluated in the generic external field
$A\equiv A(x)$. Our metric convention is $g_{\mu
  \nu}=\mathrm{diag}(-1,+1,+1,+1)$, and we use the Heaviside-Lorentz
System with $c=\hbar=1$.

\begin{figure}
\center
\includegraphics[width=0.48\textwidth]{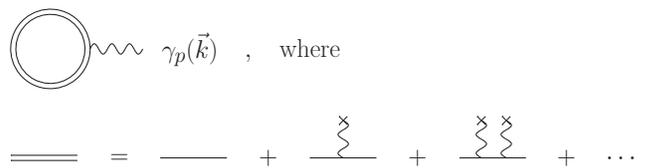}
\caption{Diagrammatic representation of the single photon {\it vacuum emission} 
process~\eqref{eq:Sp}. The double line denotes the dressed fermion propagator 
accounting for arbitrarily many couplings to the external field $\bar A$, 
represented by the wiggly lines ending at crosses.}
\label{fig:j_nonpert}
\end{figure}

In turn, the amplitude for emission of a single signal photon with momentum 
$\vec{k}$ from the QED vacuum subject to the external field $\bar A$ is given 
by \cite{Karbstein:2014fva} (cf. also Fig.~\ref{fig:j_nonpert})
\begin{equation}
 {\cal S}_{(p)}(\vec{k})\equiv\big\langle\gamma_p(\vec{k})\big| 
\Gamma_\text{int}^{(1)}[\bar A(x)] \big|0\big\rangle \,. \label{eq:Sp}
\end{equation}
Here $|\gamma_{p}(\vec{k})\rangle\equiv a^\dag_{\vec{k},p}|0\rangle$ denotes the 
single signal photon state, and $p$ labels the polarization of the emitted 
photons. Transition amplitudes to final states with more photons can be 
constructed along the same lines, but are typically suppressed because of a significantly larger phase space for the
signal photons; cf. the photon splitting process in Ref. \cite{Gies:2016czm}.
The differential number of signal photons with polarization $p$ to be measured 
far outside the interaction region is then given by
\begin{equation}
{\rm d}^3N_{(p)}(\vec{k})=\frac{{\rm d}^3k}{(2\pi)^3}\bigl|{\cal 
S}_{(p)}(\vec{k})\bigr|^2 \,. \label{eq:d3Np}
\end{equation}
Representing the photon field in Lorenz gauge as
\begin{align}
 a^{\mu}(x)&=\sum_{p}\int\frac{{\rm 
d}^3{k}}{(2\pi)^3} \frac{1}{\sqrt{2k^0}} \nonumber\\
&\quad\quad\times\Bigl(\epsilon^{\mu}_{(p)}(k)\,{\rm 
e}^{-\I kx}\,a_{\vec{k},p}+\epsilon^{*\mu}_{(p)}(k)\,{\rm 
e}^{\I kx}a^\dag_{\vec{k},p}\Bigl) ,
\end{align}
where $k^0\equiv|\vec{k}|$, $kx:=k^\mu x_\mu$ and the sum is over the two 
physical (transverse) photon polarizations, \Eqref{eq:Sp} can be expressed as
\begin{equation}
{\cal S}_{(p)}(\vec{k})=\frac{\epsilon^{*\mu}_{(p)}(k)}{\sqrt{2k^0}}\int{\rm 
d}^4x\,{\rm e}^{\I kx}\,\frac{\delta\Gamma^{1\text{-loop}}_\text{HE}[A]}{\delta 
A^\mu}\bigg|_{A=\bar A(x)} . \label{eq:Sp2}
\end{equation}

No closed-form expressions of \Eqref{eq:Sp2} for generic background
field profiles are available.  For the field configurations generated
by high-intensity lasers, which vary on length (time) scales much
larger than the Compton wavelength (time) of the electron
$\lambdabar_{\rm C}\approx3.86 \cdot 10^{-13}{\rm m}$ ($\tau_{\rm
  C}\approx1.29 \cdot 10^{-21}{\rm s}$), analytical insights are
nevertheless possible by means of a locally constant field
approximation (LCFA).

The LCFA amounts to first obtaining the Heisenberg-Euler action in
constant electromagnetic fields, $\bar F^{\mu\nu}=\partial^\mu\bar
A^\nu-\partial^\nu\bar A^\mu=\text{const.}$, resulting in a
closed-form expression $\Gamma_\text{HE}(\bar F)$. As already
determined in the original works
\cite{Heisenberg:1935qt,Schwinger:1951nm}, $\Gamma_\text{HE}(\bar F)$
is a function of the two field invariants ${\cal F}=\frac{1}{4}\bar
F_{\mu\nu}\bar F^{\mu\nu}=\frac{1}{2}(\vec{B}^2-\vec{E}^2)$ and ${\cal
  G}=\frac{1}{4}\bar F_{\mu\nu}{}^*\bar
F^{\mu\nu}=-\vec{B}\cdot\vec{E}$, where $^*\bar
F^{\mu\nu}=\frac{1}{2}\epsilon^{\mu\nu\alpha\beta}\bar
F_{\alpha\beta}$. Adopting this result for inhomogeneous 
fields, yields the LCFA approximation for the action functional,
\begin{multline}
 \Gamma_\text{HE}(\bar F)=\int{\rm d}^4x\,{\cal L}_\text{HE}(\bar F) \quad \br
\xrightarrow{\bar F\to\bar F(x)} \quad \Gamma_\text{HE}\bigl[\bar 
F(x)\bigr]=\int{\rm d}^4x\,{\cal L}_\text{HE}\bigl(\bar F(x)\bigr).
\label{eq:LCFA}
\end{multline}
Due to parity invariance of QED, the dependency of the
Heisenberg-Euler Lagrangian is actually even in $\cal G$, such that
${\cal L}_\text{HE}\bigl(\bar F)={\cal L}_\text{HE}\bigl({\cal
  F},{\cal G}^2)$ for constant fields as well as for the LCFA.
As has been argued, e.g., in Refs.~\cite{Galtsov:1982,Karbstein:2015cpa,Gies:2016yaa}, 
the deviations of the LCFA result from the corresponding exact expression for 
$\Gamma_\text{HE}$ are of order ${\cal O}\bigl((\tfrac{\upsilon}{m_e})^2\bigr)$, 
where $\upsilon$ delimits the moduli of the frequency and momentum components of 
the considered inhomogeneous field from above.

Within the LCFA, we obtain 
\cite{Karbstein:2014fva,Karbstein:2015cpa,Gies:2016yaa}
\begin{multline}
{\cal S}_{(p)}(\vec{k}) = \I
\frac{\epsilon^{*\mu}_{(p)}(k)}{\sqrt{2k^0}} \,  \int{\rm d}^4 x\, {\rm 
e}^{\I kx}\, \br
\times  \biggl[(k\bar F)_\mu \frac{\partial{\cal 
L}_\text{HE}^{1\text{-loop}}}{\partial{\cal F}}  + (k\,{}^*\!\bar F)_\mu 
\frac{\partial{\cal L}_\text{HE}^{1\text{-loop}}}{\partial{\cal G}}\biggr] , 
\label{eq:SpLCFA}
\end{multline}
where $(k\bar F)_\mu:=k^\nu\bar F_{\nu\mu}(x)$, $(k\,{}^*\!\bar 
F)_\mu:=k^\nu\,{}^*\!\bar F_{\nu\mu}(x)$ and
\begin{multline}
 \frac{\partial{\cal L}_\text{HE}^{1\text{-loop}}}{\partial{\cal F}}
 =\frac{\alpha}{2\pi}\int_{0}^{\infty}\frac{{\rm d}s}{s}\,{\rm 
e}^{-\I \frac{m_e^2}{e}s} \br
 \qquad \ \times \biggl[ \frac{ab}{a^2+b^2}\frac{as\cot(bs)}{\sinh^2(as)} 
+(a\leftrightarrow \I b)+\frac{2}{3}\biggr] , 
\end{multline} 
\begin{multline}
 \frac{\partial{\cal L}_\text{HE}^{1\text{-loop}}}{\partial{\cal G}}
 =\frac{\alpha}{2\pi}\int_{0}^{\infty}\frac{{\rm d}s}{s}\,{\rm 
e}^{-\I \frac{m_e^2}{e}s}\, {\cal G}\,\coth(as)\cot(bs)\br
 \times  \biggl[ \frac{1}{2ab} 
-\frac{1}{a^2+b^2}\frac{bs}{\sinh(as)\cosh(as)} + (a\leftrightarrow \I b) 
\biggr], \label{eq:dLdF&dLdG}
\end{multline}
with $a:=(\sqrt{{\cal F}^2(x)+{\cal G}^2(x)}-{\cal F}(x))^{1/2}$ and 
$b:=(\sqrt{{\cal F}^2(x)+{\cal G}^2(x)}+{\cal F}(x))^{1/2}$.

\begin{figure}
\center
\includegraphics[width=0.2\textwidth]{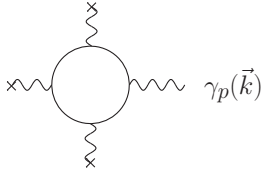}
\caption{Leading contribution to the single photon {\it vacuum emission} process 
in the limit of weak external fields.}
\label{fig:j_pert}
\end{figure}

Using spherical momentum coordinates $\vec{k}={\rm k}\hat{\vec{k}}$, where 
${\rm k}=\sqrt{k_{\rm x}^2+k_{\rm y}^2+k_{\rm z}^2}$ and 
$\hat{\vec{k}}=(\cos\varphi\sin\vartheta,\sin\varphi\sin\vartheta,
\cos\vartheta)$, the vectors perpendicular to $\vec{k}$ can be parameterized by 
a single angle $\beta$,
\begin{equation}
\hat{\vec{e}}_{\beta}=
\left(\begin{array}{c}
  \cos\varphi\cos\vartheta\cos\beta-\sin\varphi\sin\beta \\
  \sin\varphi\cos\vartheta\cos\beta+\cos\varphi\sin\beta \\
  -\sin\vartheta\cos\beta
 \end{array}\right) . \label{eq:e_perpbeta}
\end{equation}
Correspondingly, the transverse polarization modes of photons with wave vector 
$\vec{k}$ can be spanned by two orthonormalized four-vectors, e.g.,
\begin{equation}
 \epsilon^\mu_{(1)}(\vec{k}):=(0,\hat{\vec{e}}_{\beta}) \quad\textrm{and}\quad 
\epsilon^\mu_{(2)}(\vec{k}):=(0,\hat{\vec{e}}_{\beta+\frac{\pi}{2}})\,,
\label{eq:epsilons}
\end{equation}
for a suitable choice of $\beta$. With these definitions, we obtain
\begin{align}
{\cal S}_{(1)}(\vec{k})
 &=\frac{1}{\I} \sqrt{\frac{k^0}{2}} \int{\rm d}^4 x\, {\rm e}^{\I kx} \nonumber\\ 
&\quad\times\biggl\{\bigl[\hat{\vec{e}}_{\beta}\cdot\vec{E}(x)-\hat{\vec{e}}_{
\beta+\frac{\pi}{2}}\cdot\vec{B}(x)\bigr] \frac{\partial{\cal 
L}_\text{HE}^{1\text{-loop}}}{\partial{\cal F}}  \nonumber\\
&\quad\ \ \,+\bigl[\hat{\vec{e}}_{\beta}\cdot\vec{B}(x)+\hat{\vec{e}}_{\beta+\frac{\pi}{2}}
\cdot\vec{E}(x)\bigr] \frac{\partial{\cal 
L}_\text{HE}^{1\text{-loop}}}{\partial{\cal G}}\biggr\}  \label{eq:S1}
\end{align}
and ${\cal S}_{(2)}(\vec{k})={\cal
  S}_{(1)}(\vec{k})\big|_{\beta\to\beta+\frac{\pi}{2}}$, using
$\hat{\vec{e}}_{\beta+\pi}=-\hat{\vec{e}}_{\beta}$.  In the limit of
weak electromagnetic fields, $e\bar F^{\mu\nu} \ll m_e^2$,
\Eqref{eq:dLdF&dLdG} results in
\begin{equation}
\left\{
\begin{array}{c}
\dfrac{\partial{\cal L}_\text{HE}^{1\text{-loop}}}{\partial{\cal F}}\\[2ex]
\dfrac{\partial{\cal L}_\text{HE}^{1\text{-loop}}}{\partial{\cal G}}
\end{array}
\right\}
= \frac{\alpha}{\pi}\frac{1}{45}\Bigl(\frac{e}{m_e^2}\Bigr)^2
\left\{
\begin{array}{c}
4{\cal F}(x)\\[1ex]
7{\cal G}(x)
\end{array}
\right\}
+{\cal O}\bigl((\tfrac{e\bar F}{m_e^2})^4\bigr)\,,
\end{equation} 
such that \Eqref{eq:S1} becomes
\begin{align}
{\cal S}_{(1)}(\vec{k}) = &
\frac{1}{\I}\frac{e}{4\pi^2}\frac{m_e^2}{45}\sqrt{\frac{k^0}{2}}\Bigl(\frac{e}{m_e^2}\Bigr)^3 
\int{\rm d}^4 x\, {\rm e}^{\I kx} \nonumber\\
 \quad&\times \Bigl\{\,4\bigl[\hat{\vec{e}}_{\beta}\cdot\vec{E}(x)-\hat{\vec{e}}_{
\beta+\frac{\pi}{2}}\cdot\vec{B}(x)\bigr]{\cal F}(x)  \nonumber\\
&\quad+7\bigl[\hat{\vec{e}}_{\beta}\cdot\vec{B}(x)+\hat{\vec{e}}_{\beta+\frac{\pi}{2}}
\cdot\vec{E}(x)\bigr]{\cal G}(x) \Bigr\} \,,
 \label{eq:S1pert}
\end{align}
where we neglected higher-order terms of ${\cal O}\bigl((\frac{e\bar F}{m_e^2})^5\bigr)$.
The corresponding Feynman diagram is depicted in Fig.~\ref{fig:j_pert}.
Because of Furry's theorem, in QED the total number of couplings of fermion 
loops to electromagnetic fields (i.e., including the signal photon) is always 
even. For single signal photon emission, the number of couplings to the 
external field is odd.

In spherical coordinates, the differential number of signal photons of \Eqref{eq:d3Np} can finally be expressed as
\begin{equation}
{\rm d}^3N_{(p)}(\vec{k})={\rm dk}\,{\rm d}\varphi\,{\rm d}\!\cos\vartheta\,\frac{1}{(2\pi)^3}\bigl|{\rm k}{\cal 
S}_{(p)}(\vec{k})\bigr|^2 \,. \label{eq:d3Np_polarcoords}
\end{equation}
Moreover, it is convenient to introduce the total number density of
induced signal photons polarized in mode $p$ and emitted in the
direction ($\varphi$, $\vartheta$) as follows
\cite{Karbstein:2014fva},
\begin{equation}
 \rho_{(p)}(\varphi,\vartheta):=\frac{1}{(2\pi)^3}\int_0^\infty{\rm dk}\,\bigl|{\rm k}{\cal 
S}_{(p)}(\vec{k})\bigr|^2\,. \label{eq:rho}
\end{equation}
The total number of signal photons of polarization $p$ is then
obtained as $N_{(p)}:=\int_0^{2\pi}{\rm d}\varphi\int_{-1}^1{\rm
  d}\!\cos\vartheta\,\rho_{(p)}(\varphi,\vartheta)$.  Accordingly, the
total number of signal photons of any polarization is given by
$N:=\sum_{p=1}^2 N_{(p)}$, and the associated number
density by $\rho:=\sum_{p=1}^2 \rho_{(p)}$.

\section{Collision of two high-intensity laser pulses} \label{sec:2hilpcollision}

In the present work, we consider the collision of two high-intensity
laser pulses as a concrete example for our computational scheme. On
the one hand, this configuration already features a high degree of
complexity due to a substantial set of experimentally tunable laser
and geometry parameters. On the other hand, this case is sufficiently
simple to allow for analytically or semi-analytically insights which
are essential for reliably benchmarking our numerical procedure.

Let us thus assume the background electric and magnetic fields to be
generated by the superposition of two linearly polarized laser beams.
In leading-order paraxial approximation, each of these laser beams is
characterized by a single, globally fixed wave vector and its electric
and magnetic fields.  We define the normalized wave vectors of the two
laser beams $b\in\{1,2\}$ as $\hat\kappa_b^\mu=(1,\hat{\vec{e}}_{\kappa_b})$.  The
associated electric and magnetic fields are characterized by an
overall amplitude profile ${\cal E}_b$ and point in
$\hat{\vec{e}}_{E_b}$ and $\hat{\vec{e}}_{B_b}$ directions. These unit vectors are
independent of $x$ for linear polarization.  They fulfill
$\hat{\vec{e}}_{E_b}\cdot\hat{\vec{e}}_{B_b}=\hat{\vec{e}}_{E_b}\cdot\hat{\vec{e}}
_{\kappa_b}=\hat{\vec{e}}_{B_b}\cdot\hat{\vec{e}}_{\kappa_b}=0$ and
$\hat{\vec{e}}_{E_b}\times\hat{\vec{e}}_{B_b}=\hat{\vec{e}}_{\kappa_b}$.
Hence, in this case \Eqref{eq:S1pert} can be expressed as
\begin{align}
&{\cal S}_{(1)}(\vec{k})= 
\frac{1}{\I} \frac{e}{4\pi^2}\frac{m_e^2}{45}\sqrt{\frac{k^0}{2}}\Bigl(\frac{e}{m_e^2}\Bigr)^3 
\int{\rm d}^4 x\, {\rm e}^{\I kx}\, {\cal E}_1^2(x){\cal E}_2(x) \nonumber\\
&\quad\times\Bigl[4\bigl(\hat{\vec e}_\beta\cdot\hat{\vec 
e}_{E_1}-\hat{\vec e}_{\beta+\frac{\pi}{2}}\cdot\hat{\vec e}_{B_1}\bigr)
 \bigl(\hat{\vec e}_{B_1}\cdot\hat{\vec e}_{B_2}-\hat{\vec e}_{E_1}\cdot\hat{\vec 
e}_{E_2}\bigr) \nonumber\\
 &\quad\,\ -7\bigl(\hat{\vec e}_\beta\cdot\hat{\vec e}_{B_1}+\hat{\vec 
e}_{\beta+\frac{\pi}{2}}\cdot\hat{\vec e}_{E_1}\bigr)
 \bigl(\hat{\vec e}_{B_1}\cdot\hat{\vec e}_{E_2}+\hat{\vec e}_{E_1}\cdot\hat{\vec 
e}_{B_2}\bigr)\Bigr] \nonumber\\
 &\quad\quad\quad\quad+({\cal E}_1\leftrightarrow{\cal E}_2,\hat{\vec e}_{B_1}\leftrightarrow\hat{\vec 
e}_{B_2},\hat{\vec e}_{E_1}\leftrightarrow\hat{\vec e}_{E_2})\,.
 \label{eq:S1pert2laser}
\end{align}
The generalization of \Eqref{eq:S1pert2laser} to background fields 
generated by more laser beams is straightforward.
Without loss of generality we assume the beam axes of the two lasers to be confined to the xz-plane and parameterize the unit wave and field vectors as
\begin{align}
 \hat{\vec{e}}_{\kappa_b}&=\left(\begin{array}{c}
                             \sin\vartheta_b\\
                             0\\
                             \cos\vartheta_b
                            \end{array}
\right),\quad   \hat{\vec{e}}_{E_b}&=\left(\begin{array}{c}
                             \cos\vartheta_b\cos\beta_b\\
                             \sin\beta_b\\
                             -\sin\vartheta_b\cos\beta_b
                            \end{array}
\right),
\label{eq:Prop}
\end{align}
and $\hat{\vec{e}}_{B_b}=\hat{\vec{e}}_{E_b}\big|_{\beta_b\to\beta_b+\frac{\pi}{2}}$,
where the choice of $\beta_b$ fixes the polarization of the beam. 
Throughout this article, we assume $\vartheta_1=0$, such that the first laser beam propagates along the positive $\rm z$ axis.
In turn, the angle $\vartheta_2$ parameterizes the tilt of the beam axis of the second laser beam with respect to the first.
With these definitions, the terms written explicitly in \Eqref{eq:S1pert2laser} 
can be expressed as
\begin{align}
&{\cal S}_{(1)}(\vec{k})=\I \frac{\sqrt{\alpha}}{(2\pi)^{3/2}}\frac{m_e^2}{45} 
\Bigl(\frac{e}{m_e^2}\Bigr)^3(1-\cos\vartheta_2)\sqrt{\rm k} \nonumber\\
&\times\biggl\{ {\cal I}_{21}(k) (1-\cos\vartheta)
f(\beta_1+\beta_2,\beta+\beta_1-\varphi) \nonumber\\
&\quad+{\cal I}_{12}(k) \Bigl[ 
\bigl[
(1-\cos\vartheta\cos\vartheta_2)\cos\varphi-\sin\vartheta\sin\vartheta_2
\bigr] \nonumber\\
&\hspace*{3cm}\times f(\beta_1+\beta_2,\beta+\beta_2) \nonumber\\
&\quad\quad-\sin\varphi(\cos\vartheta-\cos\vartheta_2)\, 
g(\beta_1+\beta_2,\beta+\beta_2)
 \Bigl]
 \biggr\} \, , \label{eq:Sfinal}
\end{align}
where we have made use of the shorthand notations
\begin{align}
\begin{split}
 f(\mu,\nu)&:=4\cos\mu\cos\nu+7\sin\mu\sin\nu \,, \\
 g(\mu,\nu)&:=4\cos\mu\sin\nu-7\sin\mu\cos\nu \,,
\end{split} 
\end{align}
and 
\begin{equation}
 {\cal I}_{mn}(k):=\int{\rm d}^4 x\,{\rm e}^{\I {\rm k}(\hat{\vec 
k}\cdot\vec{x}-t)}\,{\cal E}_1^m(x){\cal E}_2^n(x)\,. \label{eq:Fint}
\end{equation}
Hence, the only remaining nontrivial task in determining the single
photon emission amplitude is to compute the Fourier transforms
\eqref{eq:Fint}.  As it is linear in ${\cal E}_1$ (${\cal E}_2$),
the contribution $\sim{\cal I}_{12}$ ($\sim{\cal I}_{21}$) in
Eq. \eqref{eq:Sfinal} can, for instance, be interpreted as signal
photons originating from the laser beam characterized by the field
profile ${\cal E}_1$ (${\cal E}_2$), which are scattered into a
different kinematic and polarization mode due to interactions with the
other laser beam described by ${\cal E}_2$ (${\cal E}_1$).

In a next step we specify the amplitude profiles ${\cal E}_b$
of the two laser beams, which we assume to be well-described by pulsed 
Gaussian laser beams of the following amplitude profile (cf., 
e.g., Refs. \cite{Siegman,Karbstein:2015cpa})
\begin{multline}
  {\cal E}_b(x)={\cal E}_{0,b}\,{\rm e}^{-\frac{({\rm 
z}_b-t_b)^2}{(\tau_b/2)^2}}\frac{w_{0,b}}{w_b({\rm z}_b)}\,{\rm e}^{-\frac{r_b^2}{w_b^2({\rm z})}} \br
 \times\cos\Bigl(\omega_b({\rm z_b}-t_b)+\tfrac{{\rm z}_b}{{\rm z}_{R,b}}\tfrac{r_b^2}{w_b^2({\rm 
z}_b)}-\arctan\tfrac{{\rm z}_b}{{\rm z}_{R,b}}+\varphi_{0,b}\Bigr)\,,
\label{eq:E(x)}
\end{multline}
with ${\rm z}_b:=\hat{\vec e}_{\kappa_b}\cdot(\vec{x}-\vec{x}_{0,b})$, $t_b:=t-t_{0,b}$ and $r_b:=\sqrt{(\vec{x}-\vec{x}_{0,b})^2-{\rm z}_b^2}$.
Here, ${\cal E}_{0,b}$ is the peak field strength, 
$\omega_{b}=\frac{2\pi}{\lambda_{b}}$ the photon energy and $\tau_{b}$ the pulse duration.
The beam is focused at $\vec{x}=\vec{x}_{0,b}$, where the peak field is reached for $t=t_{0,b}$. Its waist size is $w_{0,b}$ and its Rayleigh 
range is ${\rm z}_{R,b}=\pi w_{0,b}^2/\lambda_{b}$.
The widening of the beam's transverse extent as a function of ${\rm z}_b$ is encoded 
in the function $w_b({\rm z}_b)=w_{0,b}\sqrt{1+({\rm z}_b/{\rm z}_{R,b})^2}$, $\arctan\bigl(\tfrac{{\rm z}_b}{{\rm z}_{R,b}}\bigr)$ is the Gouy phase of the beam and $\varphi_{0,b}$ determines its phase in the focus. The total angular spread $\Theta_b$ and the radial beam divergence $\theta_b$ far from the beam waist are given by $\Theta_{b} = 2\,\theta_{b}\simeq 2\,\frac{w_{0,b}}{{\rm z}_{R,b}}$.

Without loss of generality, in the remainder of this article we will assume $x_{0,1}^\mu=(0,\vec{0})$, such that the temporal and spatial offsets of the two beams are fully controlled by $x_{0,2}^\mu=:(t_{0},\vec{x}_{0})$. \\

\begin{figure*}[t]
\begin{center}
 \includegraphics[width=\figlenN]{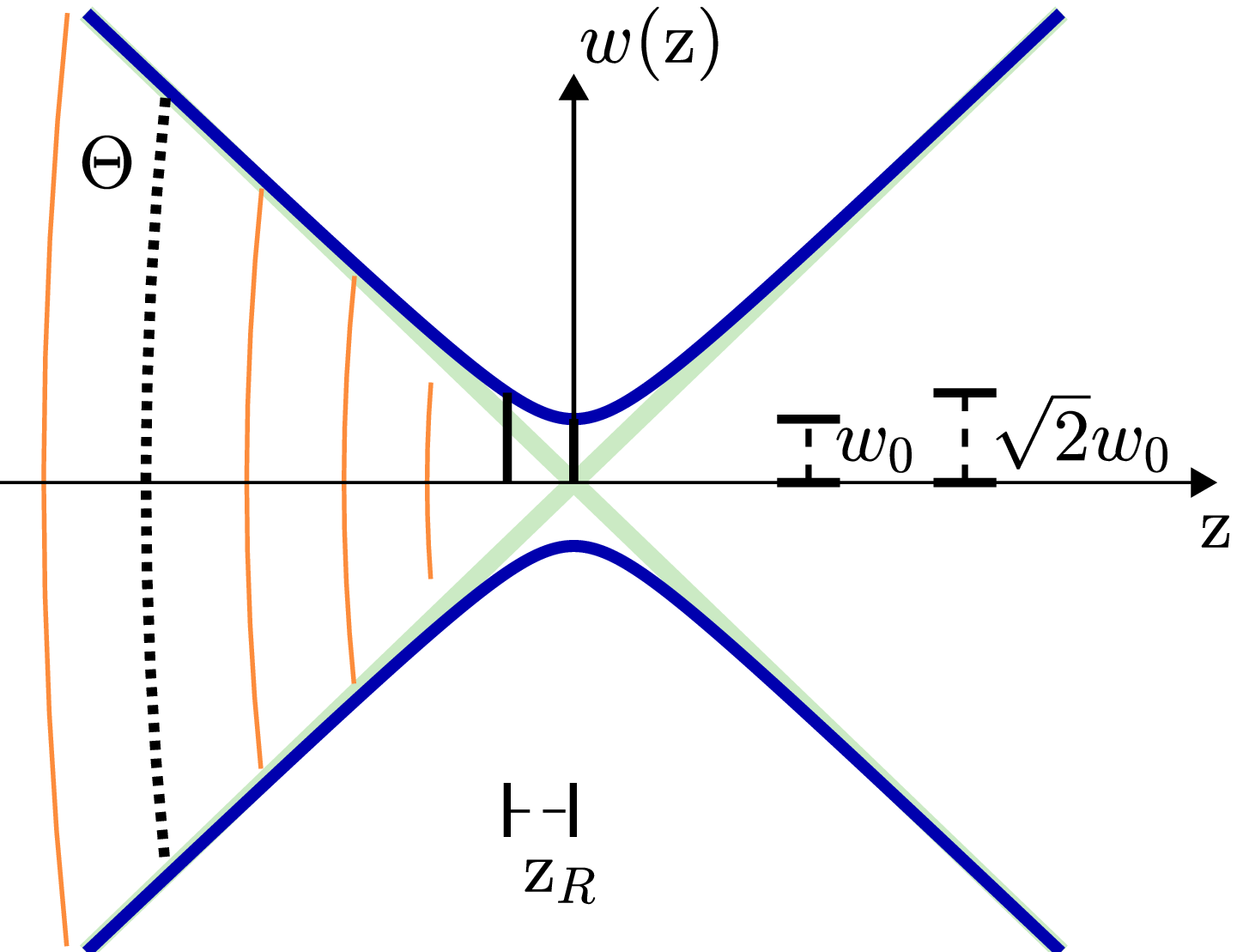}
 \includegraphics[width=\figlenN]{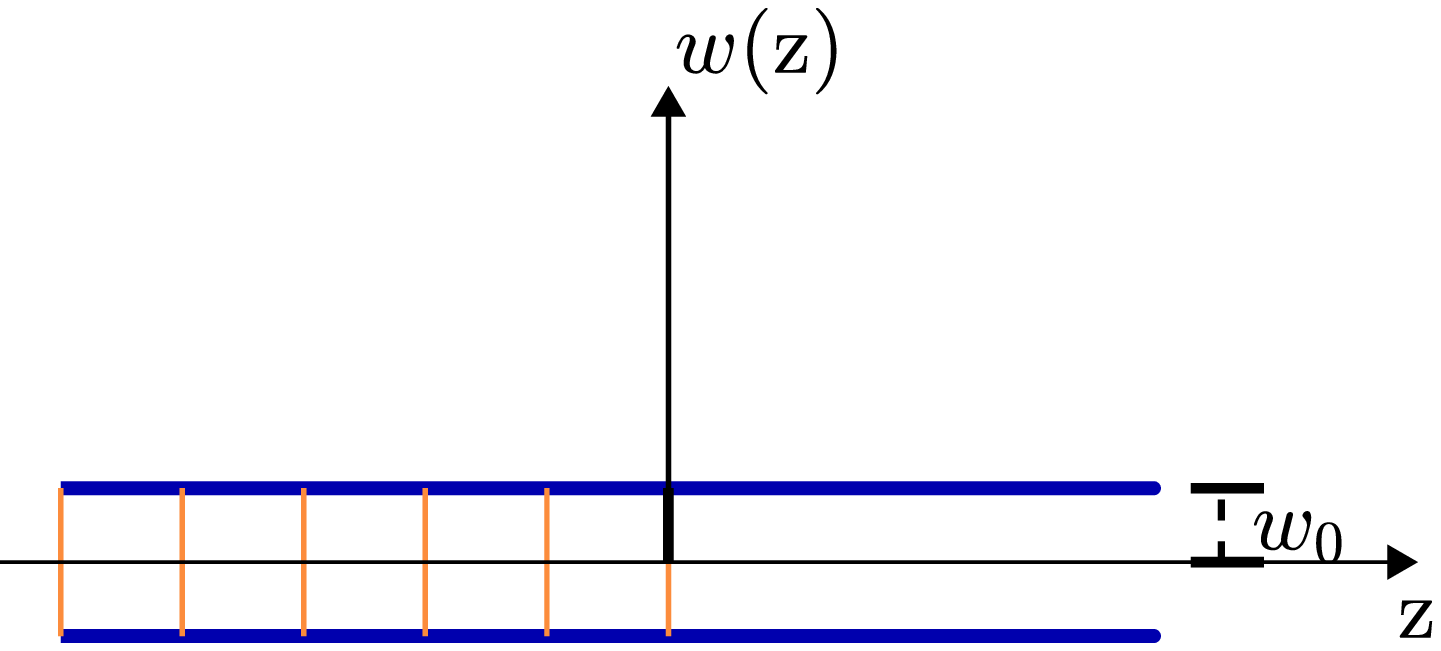}
\end{center} 
 \caption{Sketch of the transverse field amplitude profile of a generic Gaussian beam (left) and the special case of a Gaussian beam with infinite Rayleigh range, but finite beam waist (right) as a function of the longitudinal coordinate (measured along the propagation axis). 
 Here, $\Theta$ is the total angular spread, $w_{0}$ is the beam waist and ${\rm z}_{R}$ is the Rayleigh range over which the beam diameter increases by a factor of $\sqrt{2}$.}
 \label{fig:Gauss}
\end{figure*}

With regard to the Fourier integrals~\eqref{eq:Fint}, it is particularly helpful to note that the $m$-th power of the field profile~\eqref{eq:E(x)} can be expressed as
\begin{align}
 {\cal E}_b^m(x)&=\Bigl(\frac{{\cal E}_{0,b}}{2}\Bigr)^m \sum_{l=0}^m \binom{m}{l}
 c_{b;lm}({\rm z}_b,r_b) \nonumber\\
 &\quad\times{\rm e}^{\I (m-2l)[\omega_b ( {\rm z}_b -t_b ) +\varphi_{0,b}]} \, {\rm e}^{-4m \left( {\rm z}_b - t_b \right)^2/\tau_b^2}, \label{eq:Ehochm}
\end{align}
where
\begin{equation}
  c_{b;lm}({\rm z}_b,r_b)
  =\frac{{\rm e}^{-(\nicefrac{r_b}{w_{0,b}})^2 \, [\frac{m-l}{1+\I \frac{{\rm z}_b}{{\rm z}_{R,b}}}+\frac{l}{1-\I \frac{{\rm z}_b}{{\rm z}_{R,b}}}]}}{(1+\I \frac{{\rm z}_b}{{\rm z}_{R,b}})^{m-l} \, (1-\I \frac{{\rm z}_b}{{\rm z}_{R,b}})^l}\,,
\end{equation}
which can be derived straightforwardly from Eq.~(22) of Ref. \cite{Karbstein:2015cpa} by employing the binomial theorem.
Note that the entire dependence of \Eqref{eq:Ehochm} on the Rayleigh range ${\rm z}_{R,b}$ and the transverse structure of the laser fields is encoded
in the function $c_{b;lm}({\rm z}_b,r_b)$.

The integration over time in \Eqref{eq:Fint} can be easily performed
analytically for generic values of ${\rm z}_{R,b}$,
resulting in
\begin{align}
 &{\cal I}_{mn}(k)
 =\Bigl(\frac{{\cal E}_{0,1}}{2}\Bigr)^m\Bigl(\frac{{\cal E}_{0,2}}{2}\Bigr)^n \frac{\sqrt{\pi}}{2}\frac{\tau_1\tau_2}{\sqrt{m\tau_2^2+n\tau_1^2}} \nonumber\\
 &\quad\times \sum_{l=0}^m \sum_{j=0}^n \binom{m}{l} \binom{n}{j}\int{\rm d}^3x\,{\rm e}^{\I {\rm k}(\hat{\vec k}\cdot\vec{x})}\, \nonumber\\
 &\quad\times c_{1;lm}({\rm z}_1,r_1)\,c_{2;jn}({\rm z}_2,r_2) \,{\rm e}^{-4 [ m (\frac{{\rm z}_1}{\tau_1})^2 + n (\frac{{\rm z}_2+t_{0}}{\tau_2})^2] } \nonumber\\
 &\quad\times {\rm e}^{- \frac{\left\{{\rm k}+ (m-2l)\omega_1+(n-2j)\omega_2 + 8 \I [ m{\rm z}_1/\tau_1^2 + n({\rm z}_2+t_{0})/\tau_2^2] \right\}^2}{16 ( m/\tau_1^2+n/\tau_2^2 ) } } \nonumber\\
 &\quad\times{\rm e}^{\I \left\{ (m-2l)(\omega_1{\rm z}_1+\varphi_{0,1}) + (n-2j)[\omega_2 ({\rm z}_2 + t_{0}) +\varphi_{0,2}] \right\} }\,. \label{eq:Fint1}
\end{align} 
Let us now briefly focus on the limit of infinitely long pulse
durations, $\{\tau_1,\tau_2\}\to\infty$.  To this end, we first set
$\tau_2=\tau_1$ and subsequently send $\tau_1\to\infty$. This results
in the following expression,
\begin{align}
 &\lim_{\{\tau_1,\tau_2\}\to\infty}{\cal I}_{mn}(k)
 =\delta\bigl({\rm k}+(m-2l)\omega_1+(n-2j)\omega_2\bigr) \nonumber\\
 &\quad\quad\times2\pi\,\Bigl(\frac{{\cal E}_{0,1}}{2}\Bigr)^m\Bigl(\frac{{\cal E}_{0,2}}{2}\Bigr)^n  
 \sum_{l=0}^m \sum_{j=0}^n \binom{m}{l} \binom{n}{j} \nonumber\\
 &\quad\quad\times \int{\rm d}^3x\,{\rm e}^{\I {\rm k}(\hat{\vec k}\cdot\vec{x})}\,c_{1;lm}({\rm z}_1,r_1)\, c_{2;jn}({\rm z}_2,r_2) \nonumber\\
 &\quad\quad\quad\times{\rm e}^{\I \left\{ (m-2l)(\omega_1{\rm z}_1+\varphi_{0,1}) + (n-2j)[\omega_2 ({\rm z}_2 + t_{0}) +\varphi_{0,2}] \right\} } \nonumber\\
 &\quad\quad\quad\times {\rm e}^{-\I \frac{[ (m-2l)\omega_1+(n-2j)\omega_2+{\rm k}] [ m{\rm z}_1+n({\rm z}_2+t_{0}) ]}{m+n}} , \label{eq:Fint1largetaus}
\end{align}
where we have employed the identity $\lim_{\tau\to\infty} \tau\,{\rm
  e}^{-\frac{\tau^2}{2}\chi^2}=\sqrt{2\pi}\,\delta(\chi)$.  The
argument of the Dirac delta function in \Eqref{eq:Fint1largetaus}
reflects the various possibilities of energy transfer from the laser
beams to the signal photons. Due to the
strictly harmonic time dependences of the beams in the limit
$\{\tau_1,\tau_2\}\to\infty$, implying sharp laser photon energies
$\{\omega_1,\omega_2\}$, only signal photons with sharp energies ${\rm
  k}$ are induced; recall that $\{\omega_1,\omega_2,{\rm
  k}\}\geq0$. Hence, particularly for $\{\tau_1,\tau_2\}\to\infty$,
the ${\cal I}_{nm}(k)$ in \Eqref{eq:Sfinal} generically give rise to
signal photons of energy
\begin{equation}
  {\rm k}=
  \begin{cases}
      \omega_1  \\          
      \omega_1+2\omega_2  \\
      \lvert \omega_1-2\omega_2 \rvert \\
      \omega_2 \\          
      \omega_2+2\omega_1 \\
      \lvert \omega_2-2\omega_1 \rvert
  \end{cases}.
  \label{eq:feqcombs}
\end{equation}
For finite pulse durations the time dependences of the beams are no
longer purely harmonic, and correspondingly the signal frequencies in
general no longer sharp and discrete, but rather smeared and
continuous.  However, for pulse durations fulfilling
$\{\omega_1\tau_1,\omega_2\tau_2\}\gg1$, the signal frequencies
should still be strongly peaked around the values listed in
\Eqref{eq:feqcombs}.

In the limit of infinite Rayleigh ranges $\{{\rm z}_{R,1},{\rm
  z}_{R,2}\}\to\infty$, also the spatial Fourier integral in
\Eqref{eq:Fint1} can be performed analytically; cf. also
Ref. \cite{Karbstein:2016lby}. For this, we note that 
\begin{equation}
  c_{b;lm}({\rm z}_b,r_b)\,\xrightarrow{{\rm z}_{R,b}\to\infty}\,{\rm e}^{-(\nicefrac{r_b}{w_{0,b}})^2 \, m} \,. \label{eq:zRinf}
\end{equation}
Physically, the latter limit is only justified for weakly focused
laser beams, as it automatically implies $w_{0,b}\gg\lambda_b$; see the
definition of ${\rm z}_{R,b}$ in terms of $w_{0,b}$ and $\lambda_b$ given
above. In the following, we use the limit \eqref{eq:zRinf} as an
estimate also for values of $w_{0,b}/\lambda_b={\cal O}(1)$, serving
below as a toy model benchmark test for the numerical method. This
ad-hoc looking toy-model approximation can still be justified by the
following observation: The emission of signal photons from the QED
vacuum becomes substantial only in the overlap region of the focused
high-intensity laser pulses where the electromagnetic fields become
maximal.  In particular for collisions with vanishing offset of the
laser foci, the approximation based on \Eqref{eq:zRinf} is expected to
reproduce the essential quantitative features of the experimental
signal.  For an illustration of the beam profiles used, see
Fig. \ref{fig:Gauss}.

\section{Numerical implementation} \label{sec:numerics}

The vacuum emission amplitude, carrying all information about the
asymptotic signal photon, can in principle be straightforwardly
evaluated for any given external field. To the present one-loop order
within the LCFA, we may start with \Eqref{eq:SpLCFA}, or to
leading-order with \Eqref{eq:S1pert}, corresponding to a 4-dimensional
Fourier transformation from spacetime to energy-momentum space. 

In the present work, we continue to use the paraxial laser beam shapes
as an illustration. Generalizations to arbitrary spacetime-dependent
fields are straightforward on the basis of a 4-dimensional fast
Fourier transform (FFT). For the laser pulses under consideration, we
take advantage of the Gaussian time structure as in
\Eqref{eq:E(x)}. Then, the Fourier transformation in time can be
performed analytically, leaving us with a 3-dimensional space
integration (as, e.g., in \Eqref{eq:Fint1}). Reducing the integration
domain, for instance, to a cubic box, the control parameters for a
numerical integration are, e.g., the size parameter of the box
$L_{\text x},L_{\text y},L_{\text z},$ and the number of grid points
in each direction $N_{\text{x}},N_{\text{y}},N_{\text{z}}$.

The lengths $L_i$ have to be chosen large enough to enclose the
interaction region where the focused fields are strong. A natural
choice is a few times the laser focus size parameters, see App.~\ref{App} for
more details. The number of grid points is
slightly more subtle: first, this number must be high enough to
resolve the pulse structure at a sub-cycle level. Second, the grid
must also be sufficiently fine to resolve the momentum structure of
the outgoing signal photon. In the case of sum-frequency generation as
in \Eqref{eq:feqcombs}, it is this momentum scale of the signal
photons which governs the grid resolution parameters $N_i$.
Throughout this article we use a grid size of $512 \times 256 \times 512$.

Whereas the 4-dimensional integration in \Eqref{eq:SpLCFA} corresponds
to a Fourier transform, the reduced 3-dimensional case in
\Eqref{eq:Fint1} strictly speaking does not from the viewpoint of an
FFT algorithm, as the integrand also depends on the signal photon
energy ${\rm k}= \sqrt{k_{\rm x}^2+k_{\rm y}^2+k_{\rm
    z}^2}$. In practice, this is not problematic, as the integral can
still be treated as a numerical Fourier transform upon insertion of a
set of fiducial energies ${\rm k}_u$, $u=0,1, \dots, N_{\Delta({\rm
    k})}$ into the integrand. For a given ${\rm k}_u$, the
3-dimensional integral is again a Fourier transform to $k_{\rm x,y,z}$
space which we perform via FFT. The physical result then satisfies the
constraint ${\rm k}_{u} \overset{!}{=}
\sqrt{k_{\rm x}^2+k_{\rm y}^2+k_{\rm z}^2}$. In
practice, this implies that we also need to choose a grid in fiducial
${\rm k}_u$ space parametrized by a size of ${\rm k}$-grid intervals
$\Delta({\rm k})$ and the number $N_{\Delta({\rm k})}$ of intervals.
In the present case of colliding laser pulses, this discretization is
straightforward to choose as the peak locations are known from energy
conservation a la \Eqref{eq:feqcombs}, and the peak width being
inversely proportional to the pulse durations. The necessity
  of introducing a fiducial momentum grid ${\rm k}_u$ renders the
  numerical problem 4-dimensional again. Nevertheless, the advantage
  is that the spatial grid requires
  $N_{\text{x,y,z}}=\mathcal{O}(100-1000)$, whereas $ N_{\Delta({\rm
      k})}=\mathcal{O}(10)$ is sufficient for the present problem.

Concentrating on the case of colliding laser pulses as outlined above,
we observe that the spatial and directional properties of the laser
fields factorize in the general emission rate
\eqref{eq:S1pert2laser}. Thus, it is beneficial to decompose the
calculation scheme into three individual steps: (i) calculation of the
Fourier integrals ${\cal I}_{mn}$, (ii) evaluation of the factors in
\Eqref{eq:Sfinal} encoding the lasers' polarization and collision
geometry and (iii) determination of the directional emission
characteristics of the signal photons.  This specific design allows
for building highly flexible code enabling, e.g., efficient
parallelization.  For the sake of convenience, we have summarized the
scheme in Proc. \ref{alg:num}.

\begin{figure}[ht]
\begin{center}
 \includegraphics[width=0.32\textwidth]{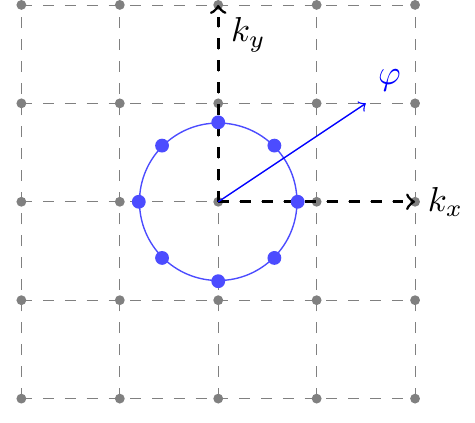}
 \hspace{3cm}
 \includegraphics[width=0.32\textwidth]{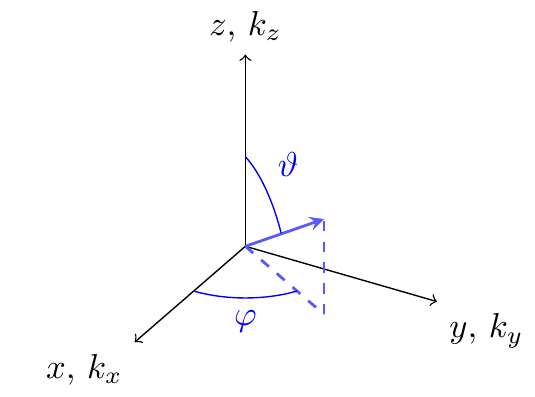}
\end{center} 
\caption{Top: Sketch of the mapping from a regular grid ($k_x$,
  $k_y$) to a polar grid with fixed radius ($\varphi$). Light gray
  (dark blue) nodes represent the discretization in Cartesian
  coordinates (polar coordinates) in momentum space. As gray and blue
  nodes generally do not overlap, we apply cubic interpolation.
  Bottom: Sketch of the coordinate
  systems used. Spatial as well as momentum coordinates are originally
  given in Cartesian coordinate systems. In spherical coordinates the
  angles $\varphi$ and $\vartheta$ give the longitude and latitude
  ($\vartheta \in \left[0, \pi \right]$), respectively. In our
  numerical calculation only regular 
  grids were used.}
 \label{fig:Num2}
\end{figure}

As the present collision set-up has a well-defined scattering center,
it is useful to characterize the signal photon in spherical momentum
coordinates $({\rm k},\varphi,\vartheta)$ rather than in Cartesian
coordinates $\left(k_{\rm x}, \, k_{\rm y}, \, k_{\rm z}
\right)$. Hence, step (i) does not only involve the FFT to $k_{\rm
  x,y,z}$ space, but also a mapping to a polar and azimuthal angle
grid discretized into $N_\vartheta$ and $N_\varphi$ intervals,
respectively. The radial momentum is already fixed by the constraint
${\rm k}= \sqrt{\vphantom{\sum} k_{\rm x}^2+k_{\rm y}^2+k_{\rm
    z}^2}$. This mapping is sketched in Fig.~\ref{fig:Num2}. 

\setlength{\extrarowheight}{3pt}
\begin{algorithm}[H]
  \caption{Pseudocode showing the general evaluation routine. The blocks are called consecutively, taking as input arguments only the results from the
  previous task.}
  \label{alg:num}
  \textbf{Code:}
   \begin{algorithmic}[50]
   \State Initialization 
   
   \ForAll{${\rm k}_u$} 
	\ForAll{${\cal I}_{mn}$}
	    \State{Fourier transform from $\left( {\rm x}, {\rm y}, {\rm z} \right)$ to $\left( k_{\rm x}, k_{\rm y}, k_{\rm z} \right)$} 
	    \State{Map from $\left( k_{\rm x}, k_{\rm y}, k_{\rm z} \right)$ to $\left( \varphi \, , \vartheta \right)$} 
	\EndFor 
    \EndFor    
   
   \ForAll{$\varphi_v \, , \vartheta_w $} 
      \State{Specify the polarization $\beta$ of the signal photons} 
      \State{Calculate emission rates $\mathcal S_{\beta}$, $\rho_{\beta}$ } 
   \EndFor
   
   \State{Post processing}

   \end{algorithmic}
\begin{tabularx}{\textwidth}{l>{$}r<{$}X}
\textbf{Notation:} &  \\
${\rm x}, \, {\rm y}, \, {\rm z}, \, k_{\rm x}, \, k_{\rm y}, \, k_{\rm z}, \, \varphi, \, \vartheta$ & \textrm{discrete variables} \\
${\rm k}_u$, $\varphi_v$, $\vartheta_w$ & \textrm{index denotes the loop variable} \\
$\left( \ldots \right)$ & \textrm{denotes a domain} \\
\end{tabularx}     
\end{algorithm}
\setlength{\extrarowheight}{0pt}

Upon combination with the functions encoding the collision geometry
and the polarization properties of the driving laser fields in
\Eqref{eq:Sfinal}, it is straightforward to obtain the discretized
version of the differential number of signal photons with energy ${\rm
  k}_u$, emitted in the direction $(\varphi_v,\vartheta_w)$ from
\Eqref{eq:d3Np}, where $u=0,\dots N_{\Delta({\rm k})},
v=1,\dots,N_\varphi, w=1,\dots,N_\vartheta$. 
Throughout this article we use $N_{\Delta({\rm k})} = 31$,  $N_\varphi = 257$ and $N_\vartheta = 513$.
Note, that at this point
the polarization properties of the signal photons have to be
specified.

The discretized version of the directional emission rate
\eqref{eq:rho} is obtained by summing over all ${\rm k}_{u}$ and is
given by
\begin{align}
 \rho_{(p)}(\varphi,\vartheta) &\approx \rho_{(p)}(\varphi_v, \vartheta_w) \nonumber\\
   &=\frac{1}{(2\pi)^3} \sum_{u=0}^{N_{\Delta ({\rm k})}} W_{{\rm k}_u}\, \vert {\rm k}_u \, \mathcal S_{(p)}({\rm k}_u, \vartheta_v, \varphi_w)\vert^2, \label{eq:rhoN}
\end{align}  
where $W_{{\rm k}_u}$ denotes a weight function that is specified by
the integration algorithm. Already simple integration routines give a
good rate of convergence. For maximum simplicity, we hence apply the
trapezoidal rule, resulting in
\begin{multline}
 \rho_{(p)}(\varphi_v, \vartheta_w) = 
    \frac{1}{(2\pi)^3} \frac{{\rm k}_{N_{\Delta ({\rm k})}} - {\rm k}_0}{2 N_{\Delta ({\rm k})}} 
    \Biggl[\vert {\rm k}_0 \, \mathcal S_{(p)}({\rm k}_0) \vert^2 \phantom{\sum_0^1} \\
      + 2 \sum_{u=1}^{N_{\Delta ({\rm k})}-1} \vert {\rm k}_u \, \mathcal S_{(p)}({\rm k}_u) \vert^2 
      + \vert {\rm k}_{N_{\Delta ({\rm k})}} \, \mathcal S_{(p)}({\rm k}_{N_{\Delta ({\rm k})}}) \vert^2\Biggr]. 
\end{multline}
The total number of signal photons polarized in mode $p$ is then
approximately given by
\begin{equation}
 N_{(p)} \approx \sum_{v=0}^{N_{\varphi}} W_{\varphi_v} \sum_{w=0}^{N_{\vartheta}} W_{\vartheta_w} \sin \left( \vartheta_w \right) \rho_{(p)} \left( \varphi_v, \vartheta_w \right) , 
\end{equation} 
with weights $W_{\varphi_v}$ and $W_{\vartheta_w}$. Similarly to
\Eqref{eq:rhoN}, even simple routines provide a good rate of
convergence. Hence, the trapezoidal rule is used again as the simplest
method.

\section{Results}\label{sec:results}

In the following, we provide explicit results for the prospective
numbers of signal photons attainable in the collision of two
high-intensity laser pulses characterized by the field profiles
introduced in Sec.~\ref{sec:2hilpcollision}.  More specifically, we
consider two identical lasers of the one petawatt (PW) class,
delivering pulses of duration $\tau=25{\rm fs}$ and energy $W=25{\rm
  J}$ at a wavelength of $\lambda=800{\rm nm}$ (photon energy
$\omega=\frac{2\pi}{\lambda} \approx 1.55{\rm eV}$).
The peak intensity of a given laser pulse in the focus is then given by \cite{Karbstein:2017jgh}
\begin{equation}
 I_{0,b}={\cal E}_{0,b}^2\approx8\sqrt{\frac{2}{\pi}}\frac{W}{\pi w_{0,b}^2\tau} \,. 
\end{equation}

As the effects of QED vacuum nonlinearities become more pronounced for higher field strengths, we aim at minimizing the beam waists $w_{0,b}$ of the driving laser beams to maximize their peak field strengths.
The minimum value of the beam waist $w_{0,b}$ is obtained when focusing the Gaussian beam down to the diffraction limit.
The actual limit is given by $w_{0,b}=\lambda_b f^\#$, where $f^\#$ is the so-called $f$-number, defined as the ratio of the focal length and the diameter of the focusing aperture \cite{Siegman};
$f$-numbers as low as $f^\#=1$ can be realized experimentally.
Being particularly interested in the maximum number of signal photons, we mainly consider the case of an optimal overlap of the colliding laser pulses and set the offset parameters $x_{0,2}^\mu=(t_{0},\vec{x}_{0})$ to zero.
Furthermore, in the remainder of this article we assume the two lasers to be polarized perpendicularly to the collision plane, corresponding to the choice of $\beta_1=\beta_2=\frac{\pi}{2}$, and to deliver pulses of the same pulse duration, $\tau_1=\tau_2=\tau$.

\subsection{Collision of laser pulses of identical frequency} \label{subsec:collA}

\begin{figure}[b]
\includegraphics[width=\figlenN]{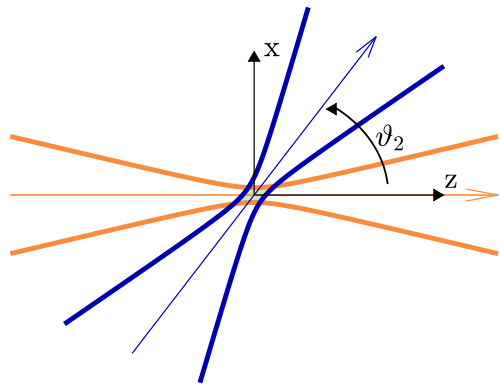}      
\caption{Sketch of the collision geometry considered in
  Sec.~\ref{subsec:collA}. Two Gaussian laser pulses collide under an
  angle $\vartheta_2$ with respect to their beam axes; the offset between
  the beam foci is $\vec{x}_{0}=0$.  Note that an angle of
  $\vartheta_2=0^\circ$($180^\circ$) corresponds to
  co(counter)-propagating laser beams.}
\label{fig:CollGeo}
\end{figure}
In a first step, we adopt the choice of $\omega_1=\omega_2=\frac{2\pi}{\lambda}$ and assume that both lasers are focused down to the diffraction limit with $f^\#=1$.
Correspondingly, we have $w_{0,1}=w_{0,2}=\lambda$. For a sketch of the considered collision geometry, see Fig.~\ref{fig:CollGeo}.
Note that the specific scenario considered here is reminiscent of the one studied in Ref.~\cite{Karbstein:2014fva}.
However, here we go substantially beyond this initial study, which only focused on exactly counter propagating beams and resorted to various additional simplifications, grasping only the most elementary features of Gaussian laser beams.

\begin{figure}[b]
\includegraphics[width=\figlenFull]{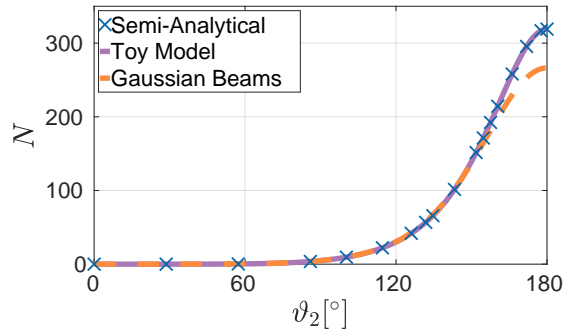}     
\caption{Total number of signal photons $N$ attainable per
  shot in the collision of two identical high-intensity laser pulses
  ($w_{0,1}=w_{0,2}=\lambda=800$nm, $W=25$J, $\tau=25$fs) plotted as a function of
  the collision angle $\vartheta_2$.  The dashed line shows the results
  for the advanced description of the colliding laser fields in terms
  of pulsed Gaussian beams, evaluated numerically with our algorithm. In addition, we
  present results for the toy-model benchmark scenario of keeping
  $w_{0,1}=w_{0,2}=\lambda$ finite but formally sending ${\rm z}_{R,b}\to\infty$. The latter scenario is analyzed in two different
  ways: By means of a fully numerical calculation with our algorithm
  (solid line), and by performing the Fourier transform from position
  to momentum space analytically, and numerically integrating over the
  outgoing signal photon momenta with Maple$^{\rm TM}$ (cross
  symbols).}
\label{fig:Ntot_An}
\end{figure}

\setlength{\extrarowheight}{3pt}
\setlength\tabcolsep{15pt}
\begin{table*}[ht]
  \caption{Benchmark calculations for the total numbers of signal photons attainable in the toy model scenario with $w_{0,1}=w_{0,2}=\lambda$ finite, but ${\rm z}_{R,b}\to\infty$; see also Figs.~\ref{fig:Ntot_An} and \ref{fig:Nperp_An}. 
    The good agreement of the results confirms the excellent performance of our numerical code. We only state the mean relative error for the total numbers of signal photons $\text{MRE}_{N}$,
    as these numbers generally show the largest deviation.}
\begin{center}
\begin{tabular}{lccccc}
 \toprule
  & \multicolumn{2}{c}{(a) Numerical} & \multicolumn{2}{c}{(b) Semi-analytical} & Mean relative error \\
  \cmidrule(lr){2-3} \cmidrule(lr){4-5}
 $\vartheta_2[{}^\circ]$ & $N$ & $N_\perp$ & $N$  & $N_\perp$ & $\text{MRE}_{N}[\%]$ \\ 
\hline
 90 & 5.03 & 0.33 & 5.04 & 0.33 & 0.2 \\ 
 135 & 69.40 & 0.59 & 69.43 & 0.60 & 0.04 \\ 
 180 & 330.19 & 0.15 & 330.24 & 0.15 & 0.02 \\
\end{tabular}
\end{center}
\label{tab:1}
\end{table*}
\setlength{\extrarowheight}{0pt}
\setlength\tabcolsep{0pt}

\begin{figure}[ht]  
\includegraphics[width=\figlenSph]{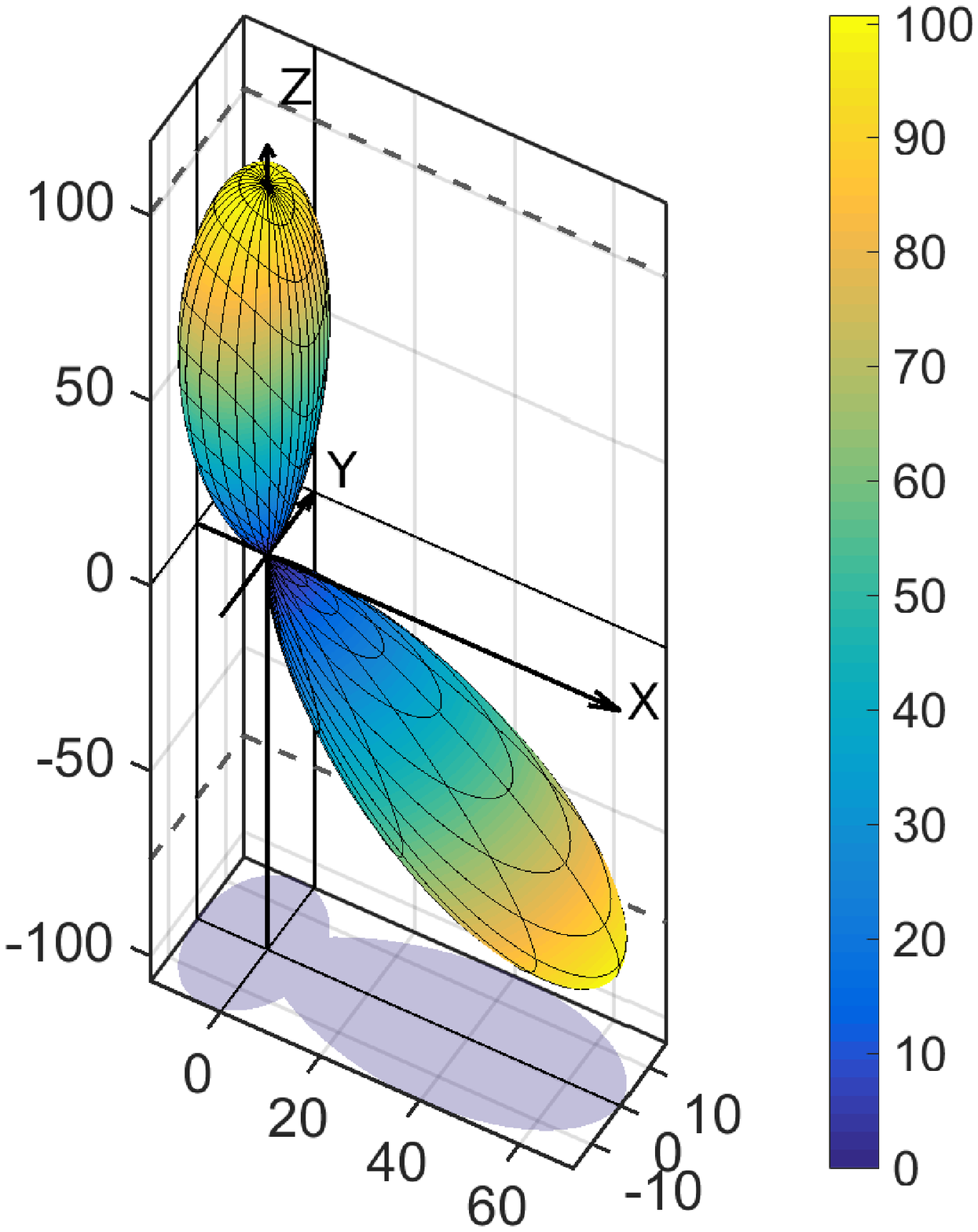}
\begin{tikzpicture}[  scale=0.95, 
        every node/.style={anchor=north west,inner sep=0pt, transform shape},
        x=20mm, y=1mm,
      ]   
     \node (fig1) at (0,0)
       {\includegraphics[trim={0cm -2cm 26cm 0cm},clip,width=\figlenSphh]{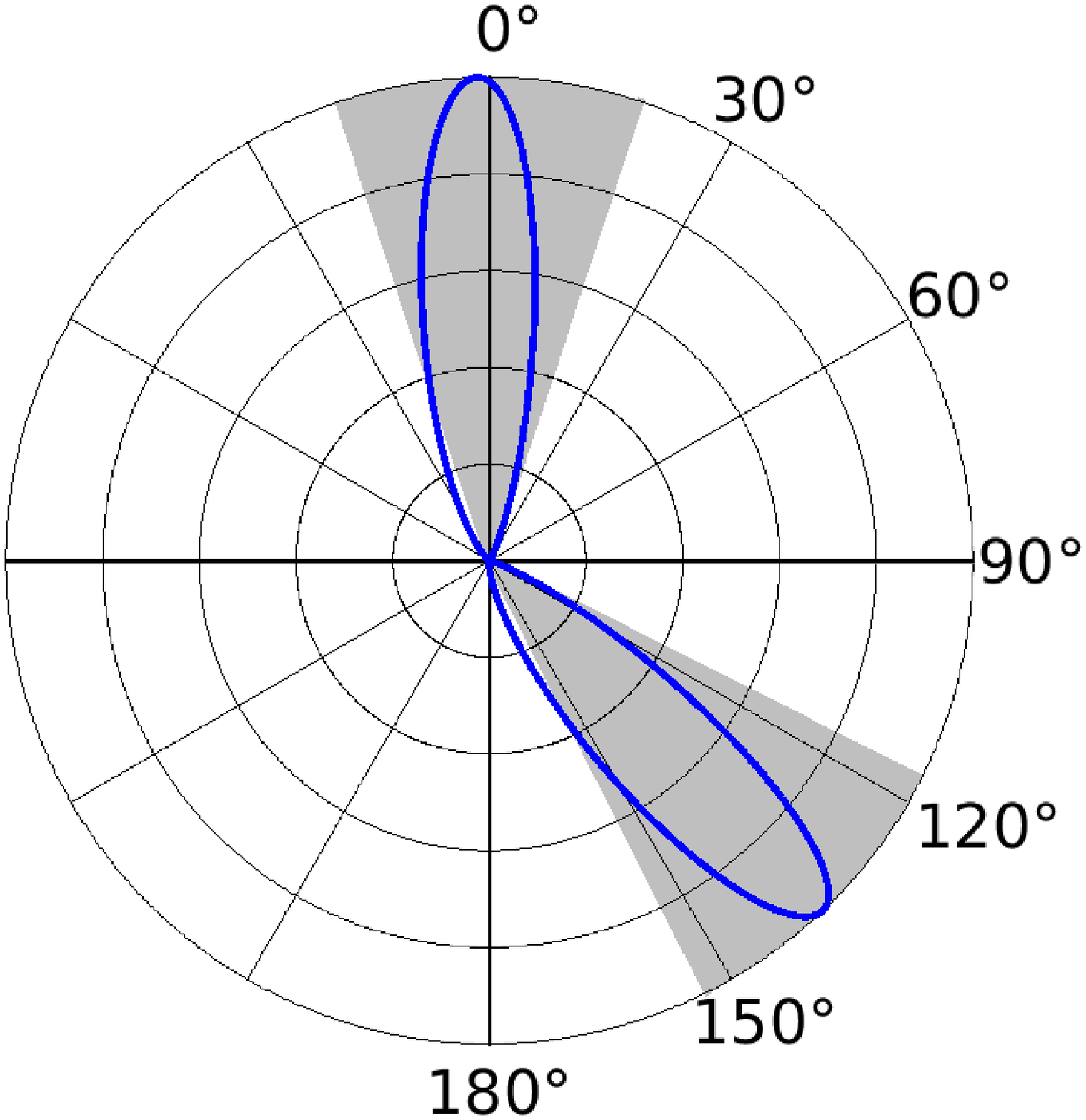} };
     \node (fig2) at (3,3)
       {\includegraphics[scale=1.1]{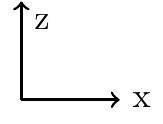}};  
\end{tikzpicture}
\vspace*{-1cm}
\caption{Directional emission characteristics of signal photons for two identical laser pulses colliding under an  angle of $\vartheta_2 = 135{}^\circ$. 
  Top: Three-dimensional
  plot of the total number density
  $\rho(\varphi,\vartheta)$.  For illustration, we also
  include a projection of the emission characteristics onto the
  xy-plane (gray). Bottom: Projection of the directional emission
  characteristics (top) onto the collision plane of the laser pulses
  (xz-plane). For comparison, the forward cones of the colliding
  Gaussian laser beams with $f^\#=1$ and delimited by the beams'
  divergences $\theta_b=\frac{1}{\pi}$ representing the background are
  highlighted in gray.}
\label{fig:EmChar1}
\end{figure}

Figure~\ref{fig:Ntot_An} shows the total number of signal photons
$N$ as a function of the collision angle $\vartheta_2$.  Here,
we depict the results for pulsed Gaussian beams with Rayleigh
ranges ${\rm z}_{R,b}$ given self-consistently by ${\rm z}_{R,b}=(\pi
  w_{0,b}^2)/\lambda=\pi\lambda$ (dashed line). We also compare it to the
toy-model benchmark scenario, where ${\rm z}_{R,b}$ is treated as an
independent parameter, which is formally sent to infinity;
cf. Sec.~\ref{sec:2hilpcollision} above.
This figure also demonstrates that the results obtained with our
numerical algorithm (solid line) for the toy-model scenario with ${\rm
  z}_{R,b}\to\infty$ are in satisfactory agreement with benchmark data
points (cross symbols).  The latter are obtained by performing the
Fourier transform from position to momentum space analytically, and
the integration over the signal photon momenta numerically using
Maple$^{\rm TM}$.  We infer that the maximum number of signal photons
is obtained for a head-on collision of the two high-intensity laser
pulses, while no signal photons are induced for co-propagating beams.
This fact is well-known from the study of probe photon propagation in
constant crossed and plane wave fields; cf., e.g.,
Ref.~\cite{Dittrich:2000zu}.  Even though for collision angles in the
range of $120^\circ\ldots180^\circ$ signal photon numbers of
$N\approx 100$ per shot are attainable, the detection
of these photons in experiment would be rather difficult.  The reason
for this is that these signal photons are predominantly emitted into
the forward cones of the incident high intensity lasers. The signal is
thus overwhelmed by the background. In Fig.~\ref{fig:EmChar1} we
exemplarily depict the directional emission characteristics for a
collision angle of $\vartheta_2 = 135{}^\circ$.  For comparison, we have
depicted the forward cones of the colliding Gaussian laser beams
focused down to $f^\#=1$ and delimited by the beams' divergences
$\theta_b=\frac{1}{\pi}$.

In order to separate a signal -- which is detectable at least in
principle -- from background, we turn to a different observable,
namely the fraction of signal photons polarized perpendicularly to the
high-intensity laser beams. Due to their distinct polarization, these
photons constitute a viable signal that could be extracted with
high-purity polarimetry.
Recall, that both high-intensity laser beams are polarized perpendicularly to the collision plane ($\beta_1=\beta_2=\frac{\pi}{2}$).

\begin{figure}[t]
\includegraphics[width=\figlenFull]{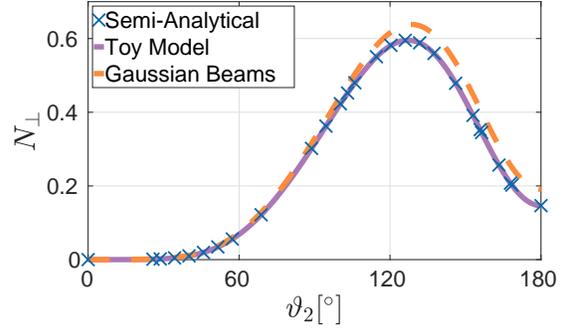}      
\caption{Total number of signal photons polarized perpendicularly to
  the high-intensity laser beams $N_\perp$ plotted as a function of
  the collision angle $\vartheta_2$.  Both laser pulses
  ($w_{0,1}=w_{0,2}=\lambda=800$nm, $W=25$J, $\tau=25$fs) are polarized
  perpendicularly to the collision plane.  The dashed (solid) curve
  shows the result obtained from a numerical calculation for pulsed
  Gaussian beams (the benchmark scenario with $w_{0,b}=\lambda$ finite,
  but ${\rm z}_{R,b} \to \infty$). The cross symbols display data for the benchmark scenario
  obtained by performing the Fourier transform analytically and
  evaluating the momentum integral numerically with Maple$^{\rm TM}$.}
\label{fig:Nperp_An}
\end{figure}

In Fig.~\ref{fig:Nperp_An} we plot the number of signal photons
polarized perpendicularly to the high-intensity laser beams $N_\perp$
as a function of $\vartheta_2$.
For the particular collision scenario considered here, this number
follows from the integration of
\begin{equation}
 \rho_{\perp}(\varphi,\vartheta):=\frac{1}{(2\pi)^3}\int_0^\infty{\rm dk}\,\bigl|{\rm k}{\cal 
S}_{(1)}(\vec{k})\bigr|^2\Big|_{\beta=-\arctan(\cos\vartheta\tan\varphi)} \label{eq:rhoperp}
\end{equation}
over the spherical angles, i.e., $N_\perp:=\int_0^{2\pi}{\rm
  d}\varphi\int_{-1}^1{\rm
  d}\!\cos\vartheta\,\rho_\perp(\varphi,\vartheta)$. Note that the
polarization-angle parameter $\beta$ has to be adjusted as a function
of the emission direction $\hat{\vec{k}}$ parameterized by
$\{\varphi,\vartheta\}$ in order to project on the perpendicular polarization
$\hat{\vec{e}}_{E_1}\cdot\hat{\vec{e}}_\beta=\hat{\vec{e}}_{E_2}\cdot\hat{\vec{e}}_\beta=0$ for all $\hat{\vec{k}}$ \cite{Karbstein:2016lby}.
As in Fig.~\ref{fig:Ntot_An}, we present results for the collision of pulsed
Gaussian laser beams, as well as for the toy model scenario with ${\rm
  z}_{R,b}\to\infty$. Again the latter scenario is used to benchmark the
performance of our numerical algorithm by comparing data points
obtained for both strategies.

For a more quantitative comparison, we exemplarily list explicit
values for the total numbers of attainable signal photons $N$ and
$N_\perp$ for several collision angles $\vartheta_2$ for the benchmark
toy-model scenario in Tab.~\ref{tab:1}.  We find a relative difference
typically on the order of $\mathcal{O}(0.01\%)$ and maximally of $\sim
0.2 \%$ between the semi-analytical approach and our numerical
algorithm. While the semi-analytical approach that involves numerical
integrations with Maple$^{\rm TM}$, we expect these algorithms to have
a higher accuracy, also because the integrations are
performed over the full (infinite) spacetime volume. The remaining
difference hence serves as an error estimate for the numerical
algorithm that works with absolute coordinate and momentum space
cutoffs due to the nature of the fast Fourier transformation.
Concretely, the fast Fourier algorithm treats the integration kernels
as if they were periodic functions.  We compensate for this by a
careful adaptation of the domain of periodicity, such that all
relevant information is preserved and no artificial frequencies are
introduced.  Additionally, the transformation to spherical coordinates
as well as the integrations over momentum space in our algorithm come
with their discretization errors. A convergence test is illustrated in
App.~\ref{App}. In summary, we consider a systematic error of our
algorithm below the $1\%$ level and thus possibly below two-loop
corrections \cite{Dittrich:1998fy} as rather satisfactory.

\begin{figure}[t]
\begin{tikzpicture}[  scale=0.95, 
        every node/.style={anchor=north west,inner sep=0pt, transform shape},
        x=10mm, y=1mm,
      ]   
     \node (fig1) at (0,0)
       {\includegraphics[width=\figlenSph]{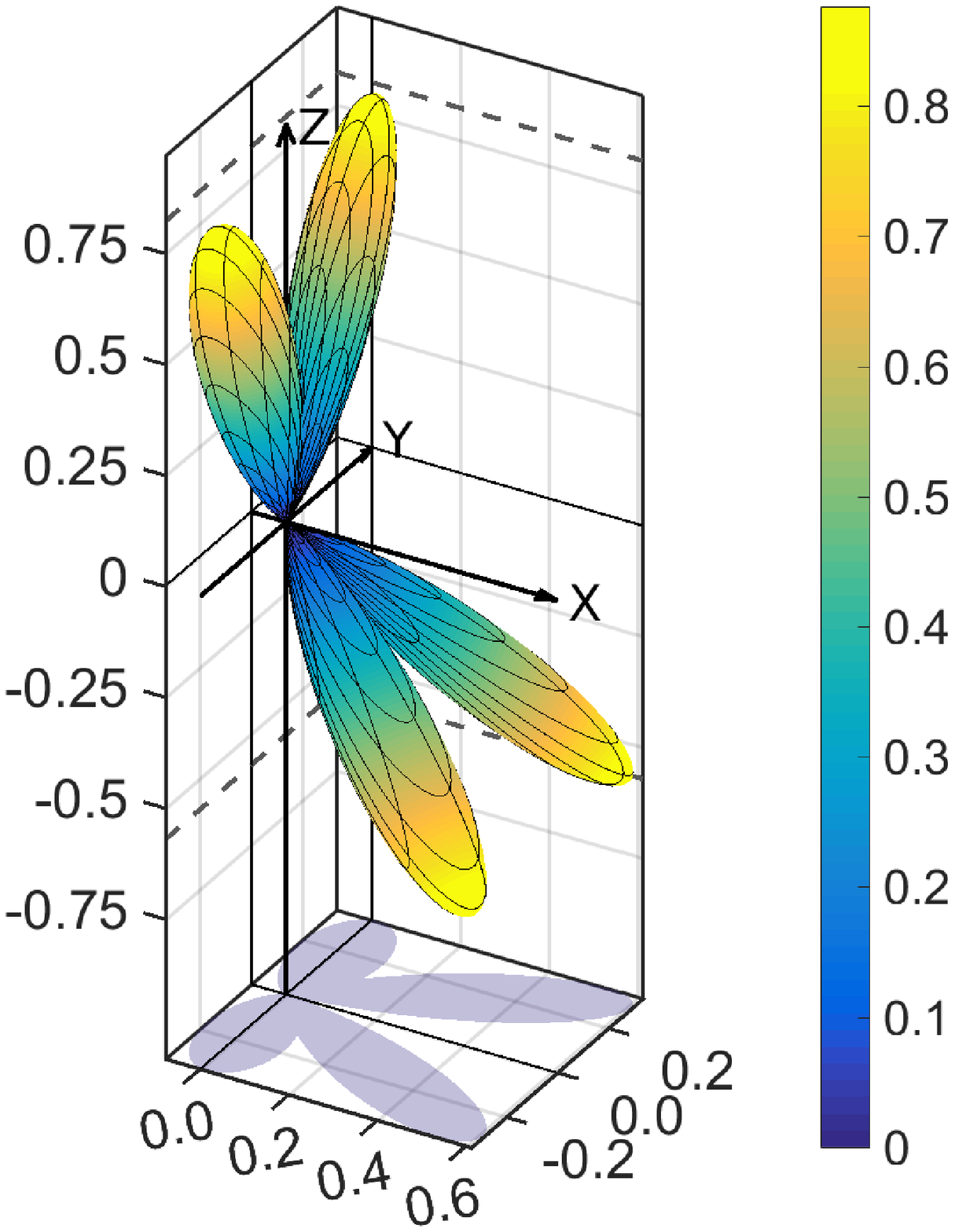}};
\end{tikzpicture}
\begin{tikzpicture}[  scale=0.95, 
        every node/.style={anchor=north west,inner sep=0pt, transform shape},
        x=20mm, y=1mm,
      ]   
     \node (fig1) at (0,0)
       {\includegraphics[trim={0cm 0cm 26cm 0cm},clip,width=\figlenSphh]{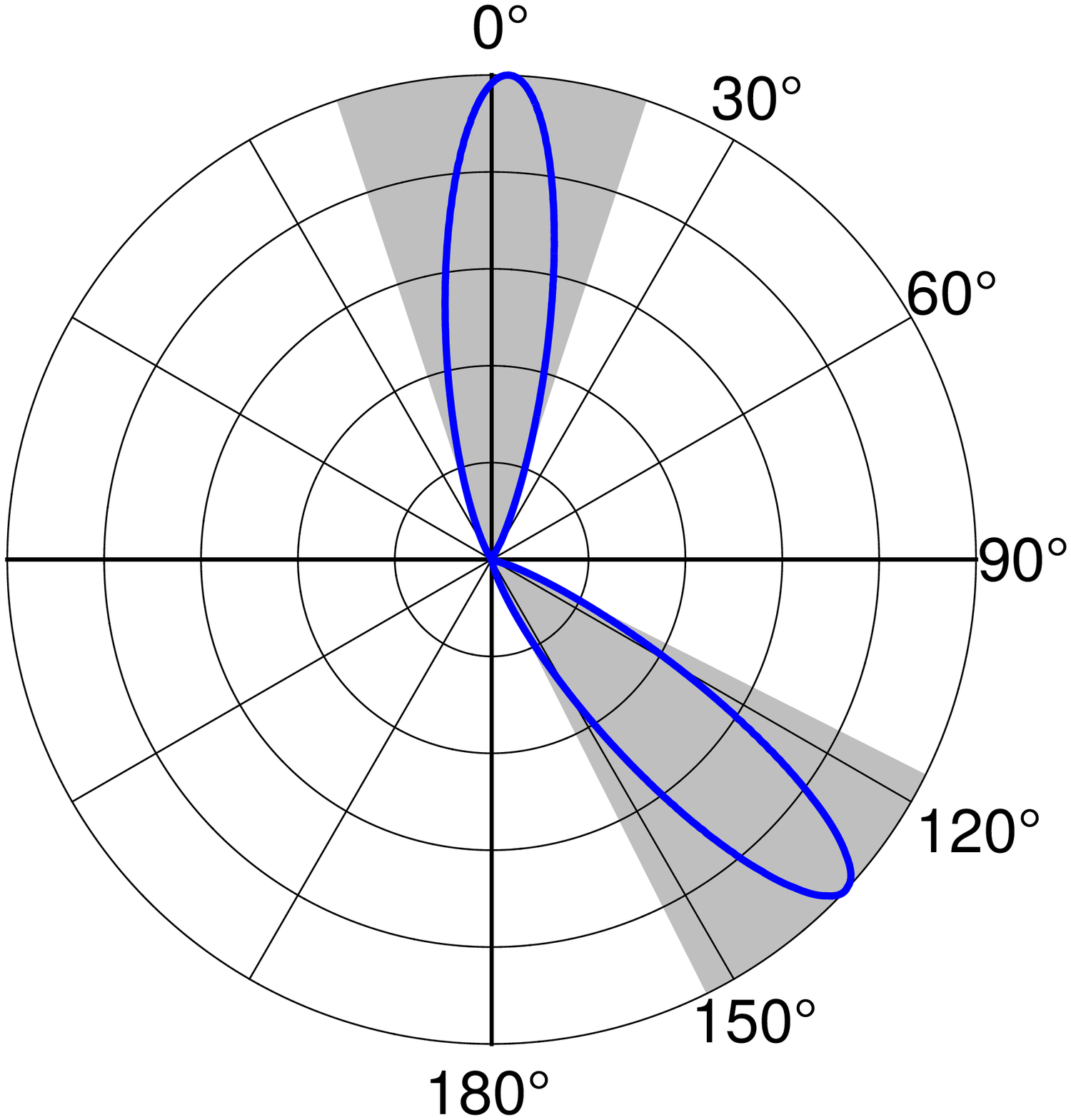} };
     \node (fig2) at (3,3)
       {\includegraphics[scale=1.1]{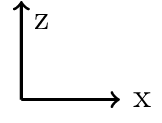}};  
\end{tikzpicture}
\vspace{-0.5cm}
\caption{Directional emission characteristics of perpendicularly
  polarized signal photons for two identical laser pulses colliding under an angle of $\vartheta_2 =
  135{}^\circ$.  Top: Three-dimensional plot of the number density
  $\rho_\perp(\varphi,\vartheta)$. Bottom: Projection of the
  directional emission characteristics (top) onto the collision plane
  of the laser pulses (xz-plane).  For comparison, the forward cones
  of the colliding Gaussian laser beams with $f^\#=1$ and delimited by
  the beams' divergences $\theta_b=\frac{1}{\pi}$ are
  highlighted in gray.}
\label{fig:Nperp_3D}
\end{figure}

\begin{figure}[t]
\begin{tikzpicture}[  scale=0.95, 
        every node/.style={anchor=north west,inner sep=0pt, transform shape},
        x=10mm, y=12mm,
      ]   
     \node (fig1) at (0,0)
       {\includegraphics[width=0.27\textwidth]{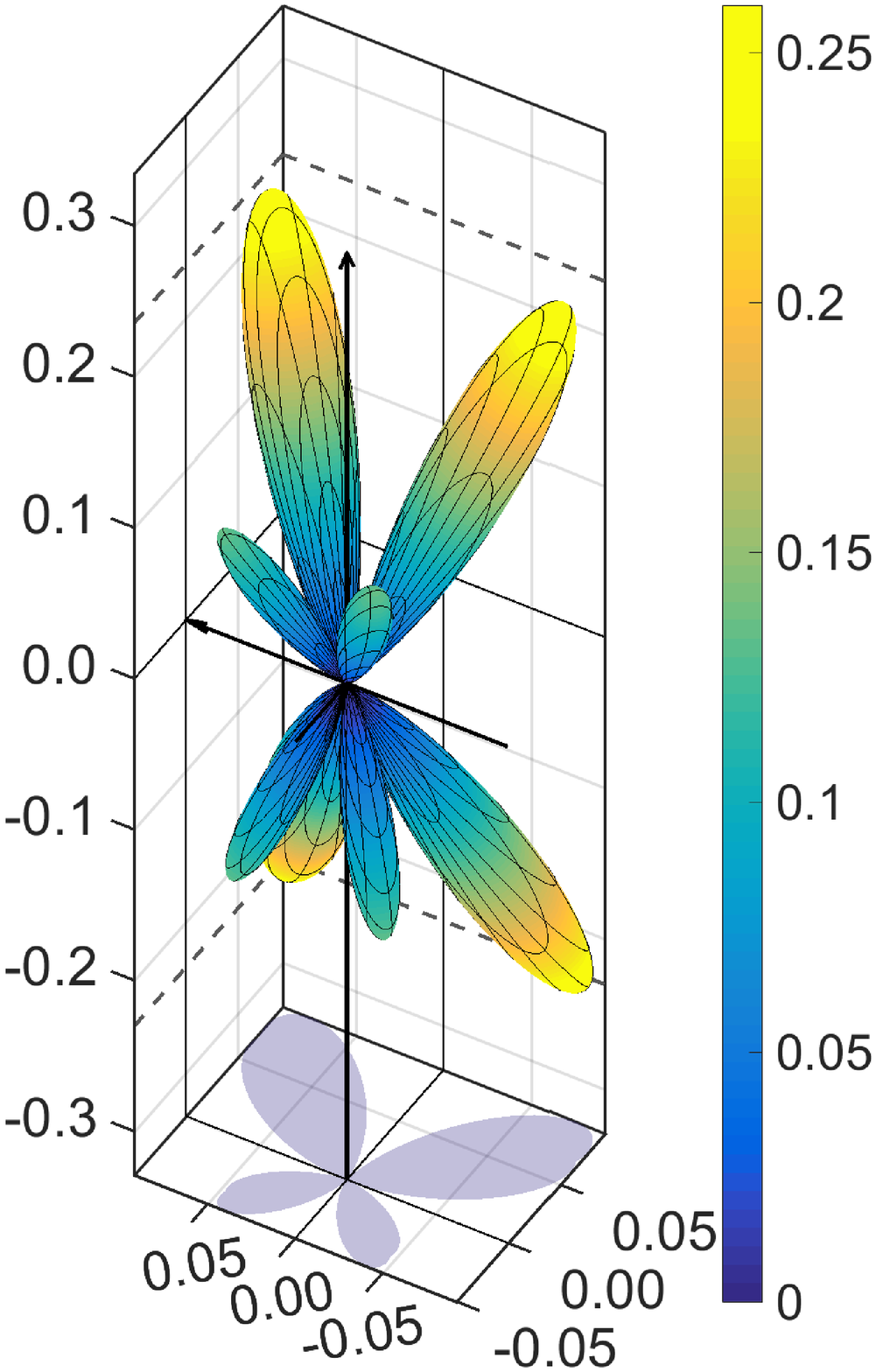}};
     \node (fig2) at (-1,1)
       {\includegraphics[scale=1.0]{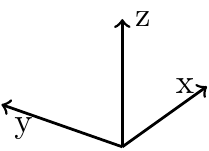}};  
\end{tikzpicture}
\begin{tikzpicture}[  scale=0.95, 
        every node/.style={anchor=north west,inner sep=0pt, transform shape},
        x=20mm, y=1mm,
      ]   
     \node (fig1) at (0,0)
       {\includegraphics[trim={0cm -2cm 26cm 0cm},clip,width=0.35\textwidth]{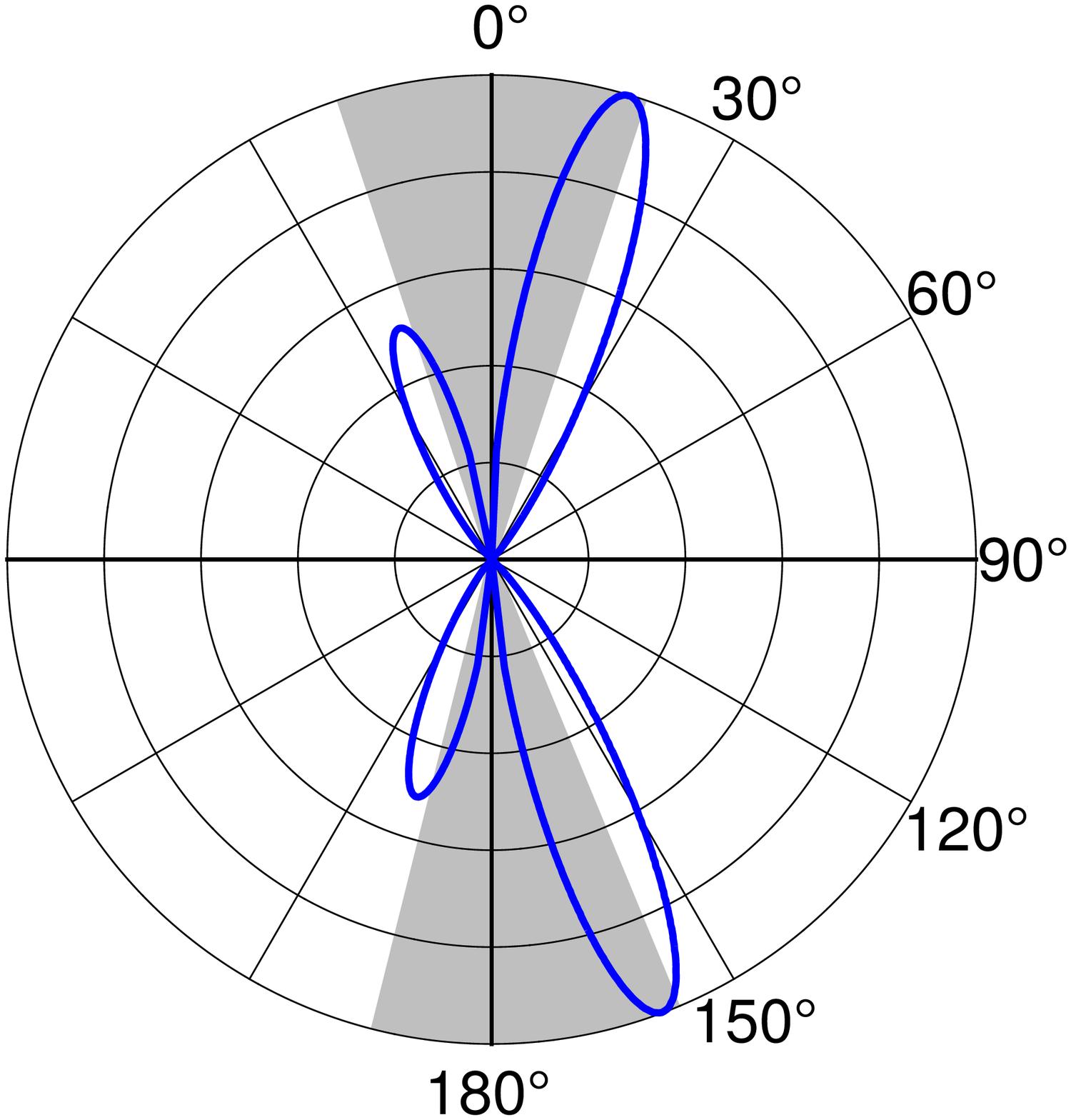} };
     \node (fig2) at (3,3)
       {\includegraphics[scale=1.1]{Fig7c}};  
\end{tikzpicture}
\vspace{-1cm}
\caption{Directional emission characteristics of perpendicularly
  polarized signal photons for two identical laser pulses colliding under an angle of $\vartheta_2 =
  175.8{}^\circ$. Top: Three-dimensional plot of the number density
  $\rho_\perp(\varphi,\vartheta)$. For better visibility of the
  directional emission characteristics, we adopt a perspective
  different from the other plots. Bottom: Projection of the
  directional emission characteristics (top) onto the collision plane
  of the laser pulses (xz-plane).  The forward cones of the colliding
  Gaussian laser beams focused down to $f^\#=1$ and delimited by the
  beams' divergences $\theta_b=\frac{1}{\pi}$ are
  highlighted in gray.}
\label{fig:Nperp_3D_2}
\end{figure}

Coming back to the physics results, Fig.~\ref{fig:Nperp_An} clearly demonstrates that the maximum for
perpendicularly polarized signal photons $N_\perp$ is shifted to a
collision angle of $\vartheta_2\approx120^\circ$.  Moreover, the
perpendicularly polarized signal is significantly smaller than the
total one; the maximum number is $N_\perp\approx0.6$.  
Analogously to Fig.~\ref{fig:EmChar1}, we also provide the
directional emission characteristics of the perpendicularly polarized
signal for a collision angle of $\vartheta_2 = 135{}^\circ$ in
Fig.~\ref{fig:Nperp_3D}.  

In addition, we display the analogous
emission characteristics for a collision angle of $\vartheta_2 =
175.8{}^\circ$ in Fig.~\ref{fig:Nperp_3D_2}.
Here, the formation of additional pronounced emission peaks opposite to the propagation directions of the high-intensity laser pulses for collision angles 
$\vartheta_2\to 180^\circ$ is clearly visible. 
For a counter-propagation geometry reflection symmetry with respect to the xy-plane is restored \cite{Karbstein:2015cpa}. 

\begin{figure}[t]
\includegraphics[width=\figlenD]{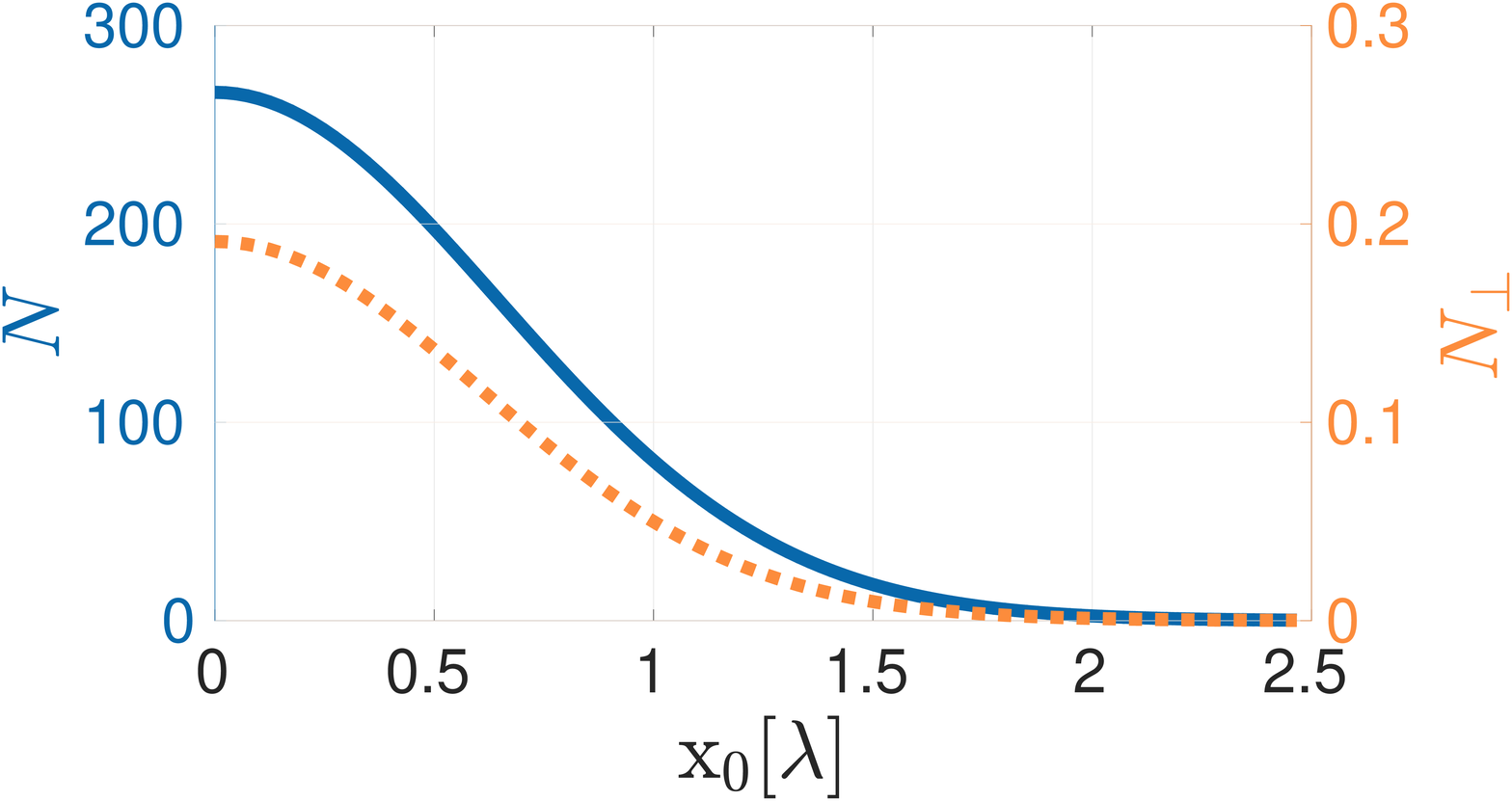}  
\hspace{1cm}
\includegraphics[width=\figlenD]{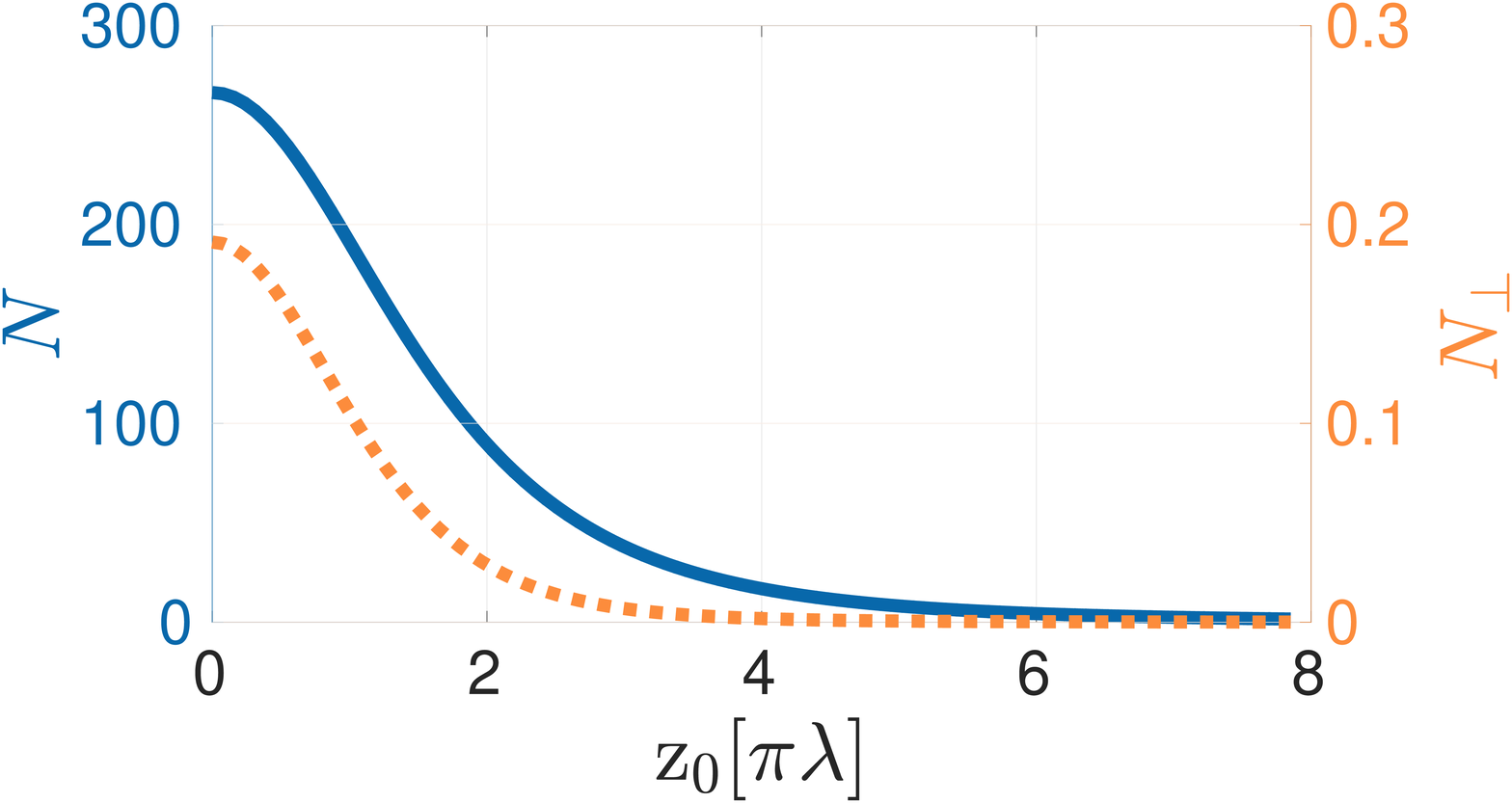}    
\caption{Impact of a relative shift between the laser foci on the
  integrated numbers of signal photons $N$ (blue solid
  line, left scale) and $N_\perp$ (orange dashed line, right scale)
  for two identical laser pulses colliding in a counter-propagation geometry, i.e., $\vartheta_2=180{}^\circ$.
  Both laser pulses are polarized perpendicularly to the collision plane.
  Top: Transverse shift with $\vec{x}_{0}=({\rm x}_{0},0,0)$ in units of
  the waist size $w_{0,1}=w_{0,2}=\lambda$. Bottom: Longitudinal shift along the common
  beam axis with $\vec{x}_{0}=(0,0,{\rm z}_{0})$ in units of the Rayleigh
  range ${\rm z}_{R,1}={\rm z}_{R,2}=\pi\lambda$.}
\label{fig:Nz0}
\end{figure}

Finally, we study the consequences of a spatial displacement $\vec{x}_{0}$
of the laser foci. Because of jitter, such a displacement is
generically expected to occur in experiments in a random fashion. For
simplicity, we specialize to the head-on collision of two identical
high-intensity laser pulses with exactly coinciding beam axes, i.e.,
$\vartheta_2=180{}^\circ$, and consider the cases $\vec{x}_{0}=({\rm
  x}_{0},0,0)$ and $\vec{x}_{0}=(0,0,{\rm z}_{0})$ focused to $w_{0,1}=w_{0,2}=\lambda$. We demonstrate in
Fig.~\ref{fig:Nz0} how the integrated numbers of signal photons
$N$ and $N_\perp$ decrease as a function of the relative
displacements ${\rm x}_{0}$ and ${\rm z}_{0}$ between the laser foci
transverse to or along the common beam axis. For the present case, we
observe that the signal photon number $N$ drops by a factor of 2 for ${\rm x}_0 \approx 0.76 \lambda$ and ${\rm z}_0 \approx 1.5 \pi\lambda$.

\subsection{Collision of laser pulses of fundamental and doubled frequency}\label{subsec:collB}

\begin{figure}[b]
\includegraphics[width=\figlenN]{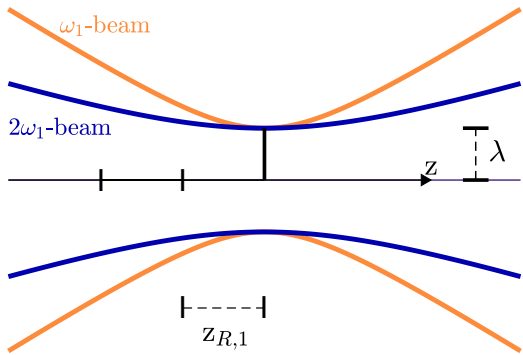}      
\caption{Scenario (i): The two Gaussian beams of fundamental and doubled frequency are focused to a beam waist of
  $w_{0,1}=w_{0,2}=\lambda$.  In this scenario, the beam divergences
  fulfill $\theta_2=\theta_1/2$. We depict the
  case of zero offset, $\vec{x}_{0}=0$, and
  $\vartheta_2\in\{0^\circ,180^\circ\}$.}
\label{fig:beams(i)}
\end{figure}

\begin{figure}[t]
\includegraphics[width=\figlenN]{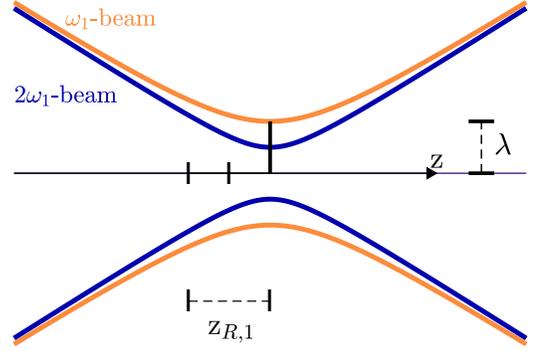}
\caption{Scenario (ii): The Gaussian beam of fundamental (doubled) frequency is
  focused to a waist size of $w_{0,1}=\lambda$ ($w_{0,2}=\lambda/2$). In this scenario, the beam divergences fulfill
  $\theta_1=\theta_2$. We depict the case of zero
  offset, $\vec{x}_0=0$, and
  $\vartheta_2\in\{0^\circ,180^\circ\}$.}
\label{fig:beams(ii)}
\end{figure}

Here we go beyond the scenario considered in the previous section,
subsequently referred to as scenario (o).  Differently to
Sec.~\ref{subsec:collA}, one of the two high-intensity lasers is now
assumed to be frequency doubled, such that $\omega_2=2\omega_1=2\frac{2\pi}{\lambda}$.
The energy loss for a frequency-doubling process conserving the pulse duration is estimated conservatively as $50\%$.
Correspondingly, we have
$\tau_1=\tau_2=\tau$, $W_1=W$ and $W_2=W/2$.  Keeping the focusing of the
fundamental-frequency laser pulse as in the previous section, i.e., $w_{0,1}=\lambda$, we now consider two
different scenarios: (i) In order to ensure a maximal spatial overlap
of the two laser pulses in their foci, the frequency-doubled laser pulse is
focused down to the waist size of the fundamental-frequency laser pulse, i.e.,
$w_{0,2}=w_{0,1}=\lambda$. This scenario is illustrated in
Fig.~\ref{fig:beams(i)}.
(ii) For maximizing the peak field strength in the focus, the
frequency-doubled pulse is focused down to its diffraction limit with
$f^\#=1$, resulting in $w_{0,2}=\lambda/2$. This scenario is
sketched in Fig.~\ref{fig:beams(ii)}.

\begin{figure}[t]
\includegraphics[width=\figlendN]{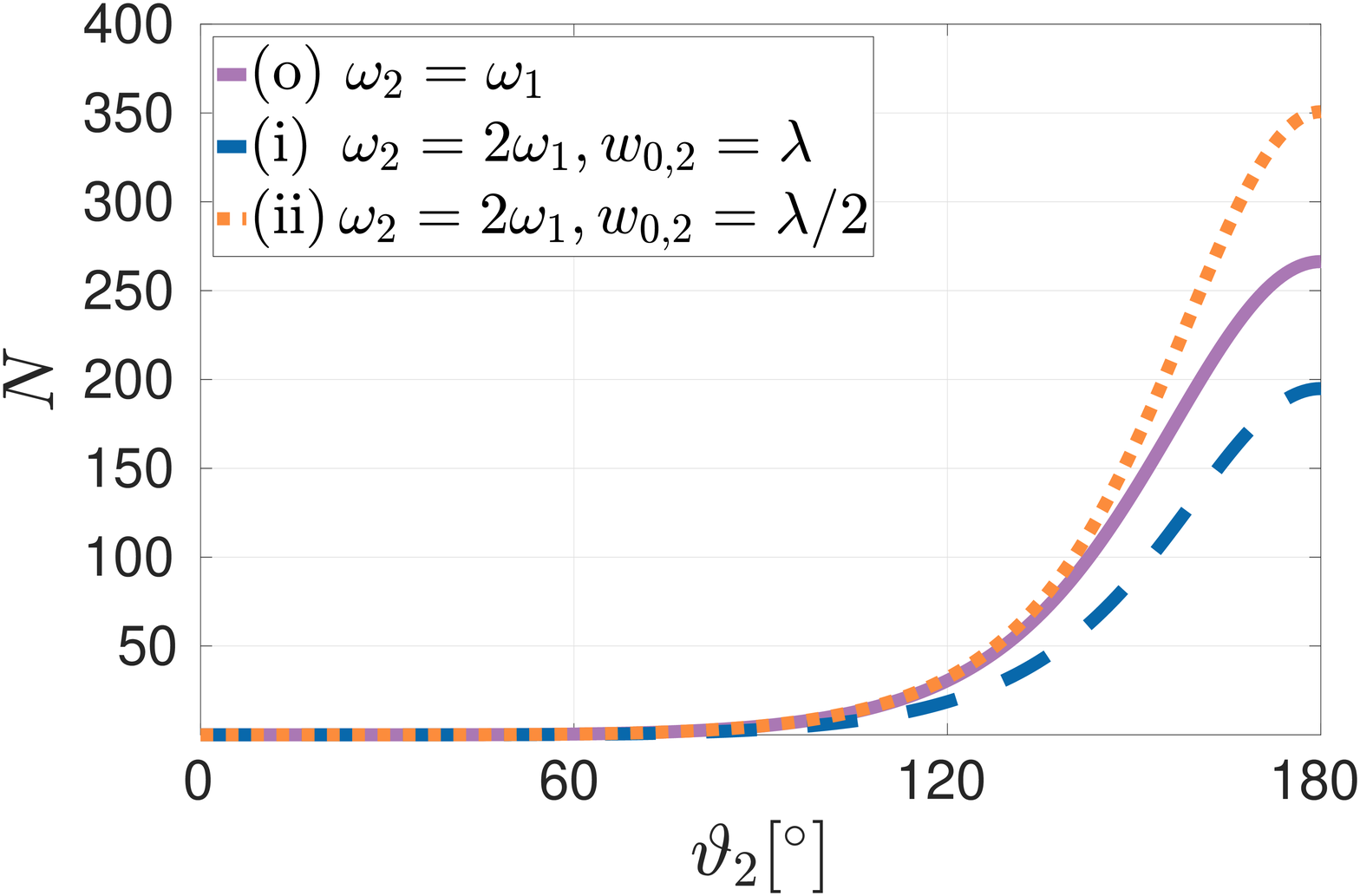}  
\includegraphics[width=\figlendN]{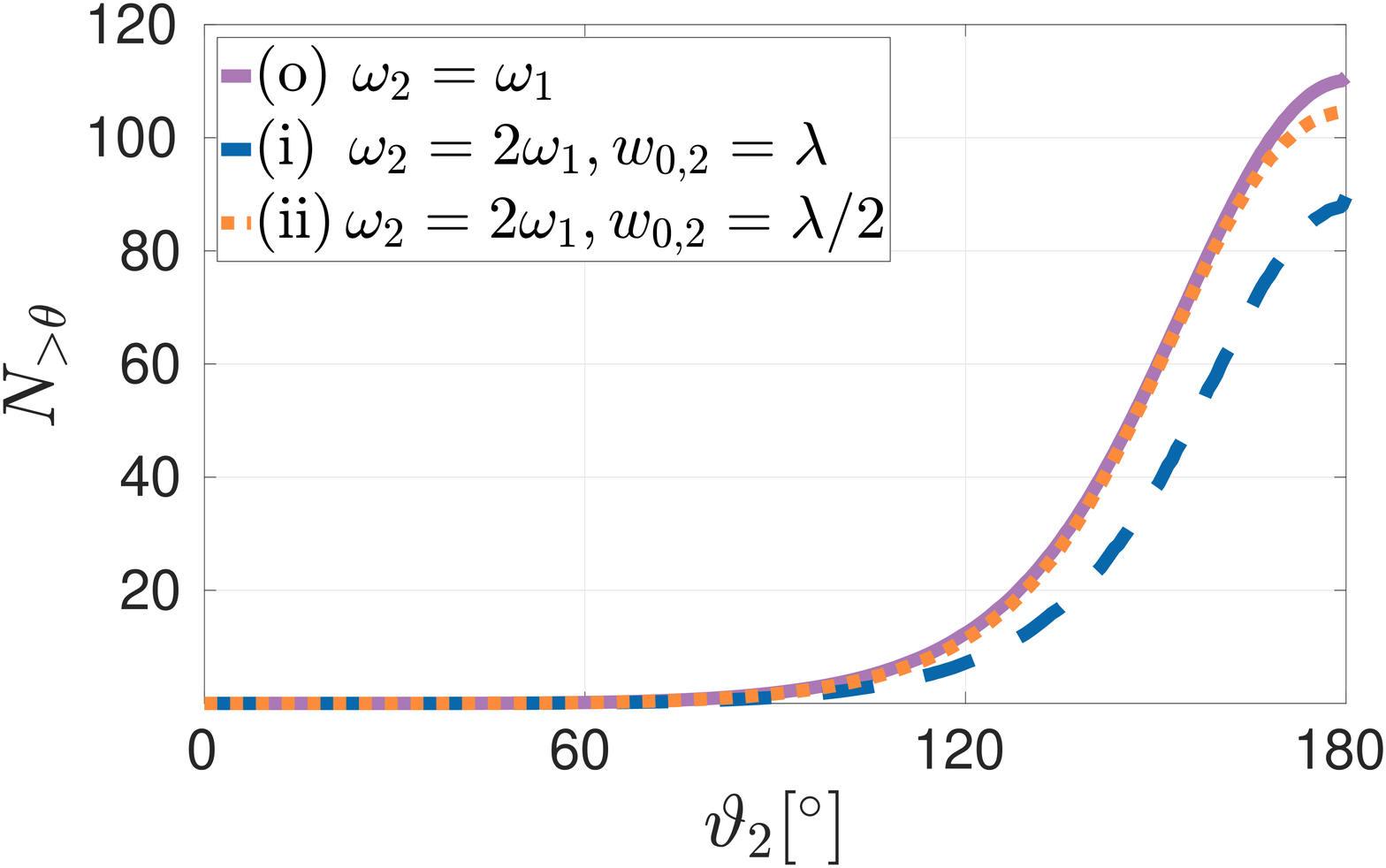}  
\caption{Integrated numbers of signal photons attainable in the
  various scenarios (o)-(ii) plotted as a function of the collision
  angle $\vartheta_2$.  We depict results for (o) the collision of two
  fundamental frequency laser pulses
  focused to $w_{0,1}=w_{0,2}=\lambda$, and the collision of
  fundamental and doubled frequency laser pulses focused to waist sizes
  (i) $w_{0,2}=w_{0,1}=\lambda$ and (ii) $w_{0,2}=w_{0,1}/2=\lambda/2$.
  Top: Total number of signal photons $N$. Bottom: Number of
  signal photons emitted outside the forward cones of the colliding
  Gaussian laser beams $N_{>\theta}$, delimited by
  the beams' radial divergences $\theta_b$.}
\label{fig:Ntot_div}
\end{figure}

As detailed in Sec.~\ref{sec:2hilpcollision}, for Gaussian beams the
Rayleigh range and far-field beam divergence, are intimately related
to the wavelength and the waist size.  Hence, in case (i) we have
${\rm z}_{R,2}=2{\rm z}_{R,1}$, $\theta_2=\theta_1/2$, while in case
(ii) ${\rm z}_{R,2}={\rm z}_{R,1}/2$, $\theta_2=\theta_1$; cf. also Figs.~\ref{fig:beams(i)} and
\ref{fig:beams(ii)}.  All the results presented in this section are
obtained with our algorithm introduced in Sec.~\ref{sec:numerics}.
In Fig.~\ref{fig:Ntot_div} we show the total number of
signal photons $N$ as a function of the collision angle
$\vartheta_2$ for the cases (o)-(ii).

In Sec.~\ref{sec:2hilpcollision} we have argued that the signal
photons should predominantly be emitted at several pronounced
frequencies if the criterion
$\{\omega_1\tau,\omega_2\tau\}\gg1$ holds;
cf.~\Eqref{eq:feqcombs}.  For the collision of (o) two fundamental
frequency beams we have $\omega_1\tau=\omega_2\tau \approx
58.9$, while for the cases (i) and (ii), both involving a frequency
doubled beam, we have $\{\omega_1\tau,\omega_2\tau\} \approx
\{58.9, 117.7 \}$.

Hence, as the criterion $\{\omega_1\tau,\omega_2\tau\}\gg1$ is
obviously fulfilled here, we expect the signal photons to feature
primarily frequencies with (o): ${\rm k}\approx\{\omega_1,3\omega_1\}$ and
(i), (ii): ${\rm k}\approx\{\omega_1,2\omega_1,3\omega_1,4\omega_1,5\omega_1\}$,
respectively.  However, inelastic signal photon emission processes are
generically suppressed in comparison to the elastic ones.  For
instance, in Ref.~\cite{Karbstein:2014fva} it was already demonstrated
for a simplified model of the head-on collision of fundamental
frequency laser pulses that the $3\omega_1$ signal is completely
negligible in comparison to the $\omega_1$ signal.  This fully agrees
with the results obtained here: In scenario (o) essentially all signal
photons are emitted in an energy range $\Delta(\omega_1)$; here and in the following
$\Delta(\omega)$ denotes an interval of photon energies centered around a
frequency $\omega$ with an energy width being inversely proportional to the
temporal pulse duration.
For the
scenarios (i) and (ii) we encounter sizable numbers of signal photons
in the energy segments $\Delta(\omega_1)$ and $\Delta(2\omega_1)$.

\setlength{\extrarowheight}{3pt}
\setlength\tabcolsep{11pt}
\begin{table*}[t]
  \caption{Prospective numbers of signal photons with energies in the segments $\Delta(\omega_1)$ 
    and $\Delta(2\omega_1)$ for the example of a collision angle of $\vartheta_2 = 135^\circ$.
    Both high-intensity laser pulses are polarized perpendicularly to the collision plane.
    Apart from (o) the collision of two identical beams of frequency $\omega_1$ focused 
    to $w_{0,1}=w_{0,2}=\lambda$, we consider collisions of fundamental-frequency $\omega$ 
    and frequency-doubled $\omega_2=2\omega_1$ beams focused to (i) $w_{0,1}=w_{0,2}=\lambda$, 
    and  (ii) $w_{0,1}=2 w_{0,2}=\lambda$.
    We provide values for the total (perpendicularly polarized) number of signal photons 
    $N$ ($N_\perp$). Besides, $n_{>\theta}$ ($n_{\perp,>\theta}$) denotes the fraction of 
    $N$ ($N_\perp$) emitted outside the forward divergence of the Gaussian high-intensity lasers.}
\begin{center}
\begin{tabular}{ccccccccc}
 \toprule
  & \multicolumn{4}{c}{$\Delta(\omega_1)$} & \multicolumn{4}{c}{$\Delta(2\omega_1)$} \\
  \cmidrule(lr){2-5} \cmidrule(lr){6-9}
 \text{scenario} & $N$ & $n_{>\theta}$ & $N_\perp$ & $n_{\perp,>\theta}$ & $N$ & $n_{>\theta}$ & $N_\perp$ & $n_{\perp,>\theta}$ \\ 
\hline
 (o) & 70.53 & 42\% & 0.66 & 74\% & - & - & - & - \\
 (i) & 9.20 & 44\% & 0.08 & 75\% & 34.24 & 40\% & 0.10 & 75\% \\
 (ii) & 24.02 & 66\% & 0.35 & 90\% & 53.67 & 24\% & 0.29 & 54\%
\end{tabular}
\end{center}
\label{tab:2}
\end{table*}
\setlength{\extrarowheight}{0pt}
\setlength\tabcolsep{0pt}

\begin{figure}[t]
\includegraphics[width=\figlendN]{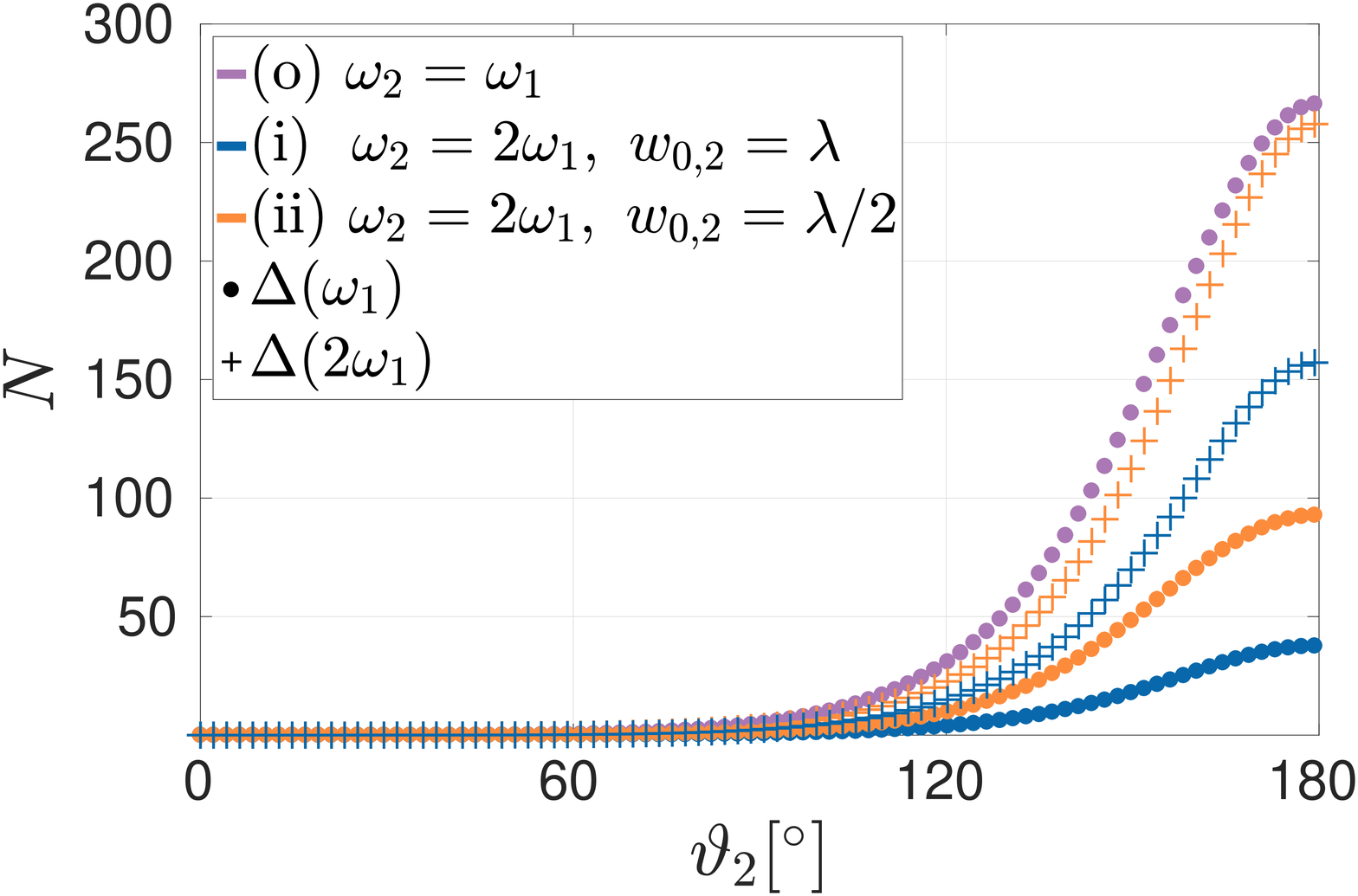}    
\includegraphics[width=\figlendN]{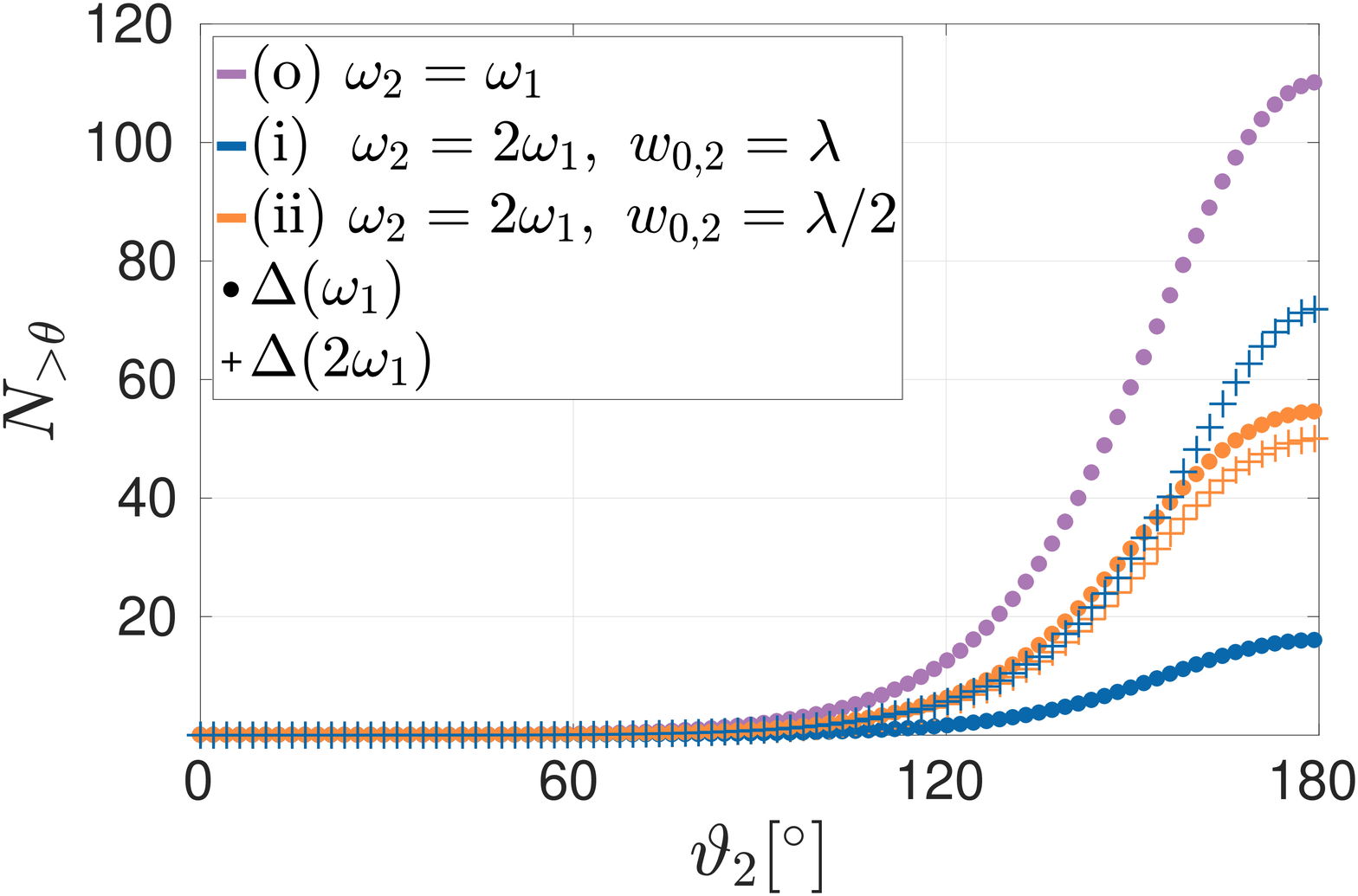}   
\caption{Partitioning of the attainable numbers of signal photons into
  the energy regimes $\Delta(\omega_1)$ and $\Delta(2\omega_1)$ for the
  various scenarios (o)-(ii). Both high-intensity laser pulses are polarized perpendicularly to the collision plane. 
  The segment with center frequency ${\rm
    k} =\omega_1$ ($2\omega_1$) is depicted by $\bullet$ ($+$) symbols.
  Naturally, there is no ${\rm k}\approx2\omega_1$ signal for the
  collision of two fundamental frequency beams. Top: Total number of
  signal photons $N$. Bottom: Integrated number of signal
  photons emitted outside the forward cones of the colliding Gaussian
  laser beams $N_{>\theta}$.}
\label{fig:Ntot_div_char}
\end{figure}

In Fig.~\ref{fig:Ntot_div_char} we show the partitioning of the
emitted signal photons into the dominant frequency channels ${\rm
  k}\approx\{\omega_1,2\omega_1\}$.  We present results for the total
number of attainable signal photons $N$ for all the
scenarios (o)-(ii) introduced above.  In addition, we provide the
number of signal photons $N_{>\theta}$ emitted outside the forward
cones (delimited by the beam divergences $\theta_b$) of the high-intensity
lasers.
\FloatBarrier
\begin{figure}[t]
\includegraphics[width=\figlenCut]{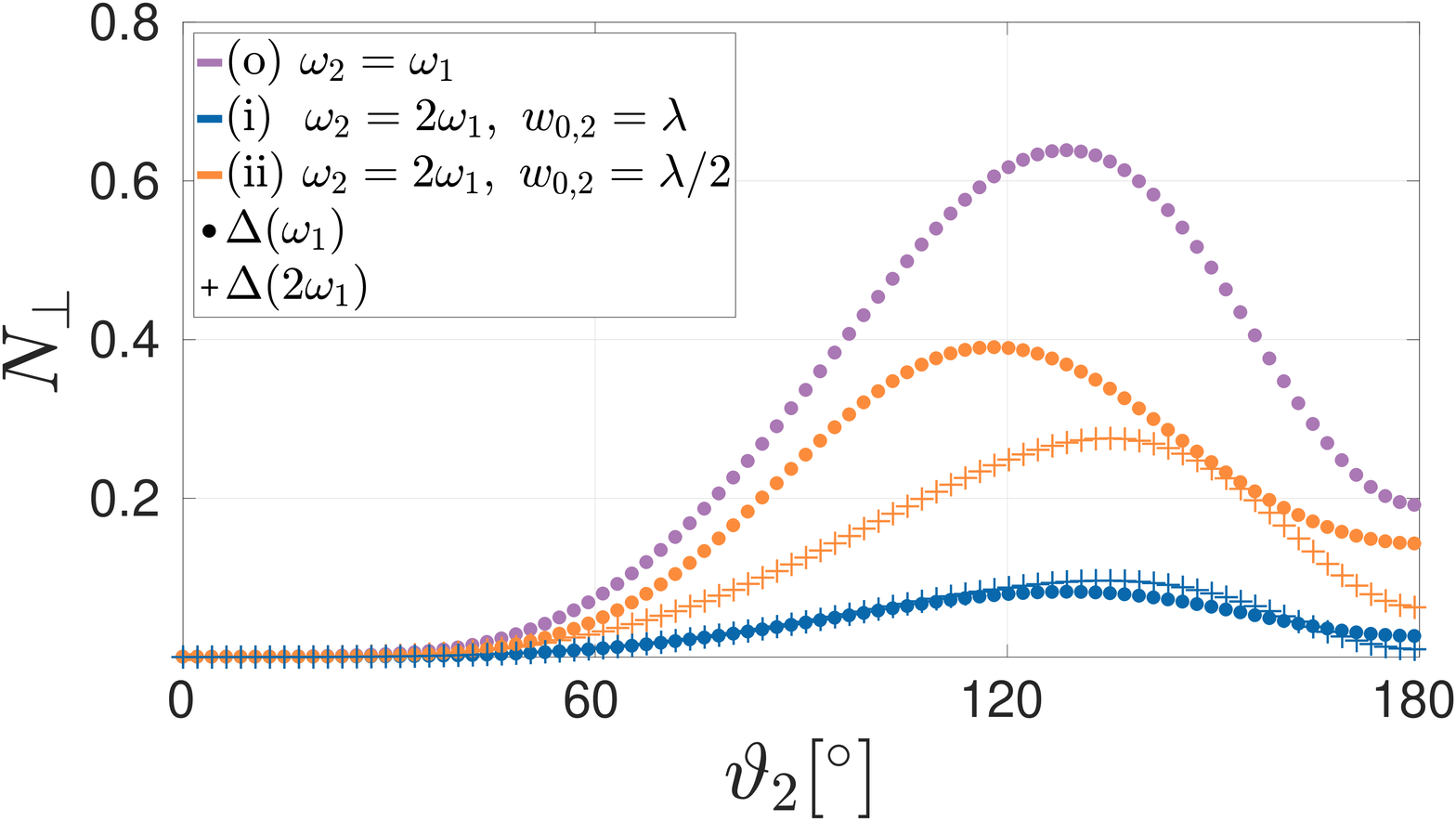}    
\includegraphics[width=\figlenCut]{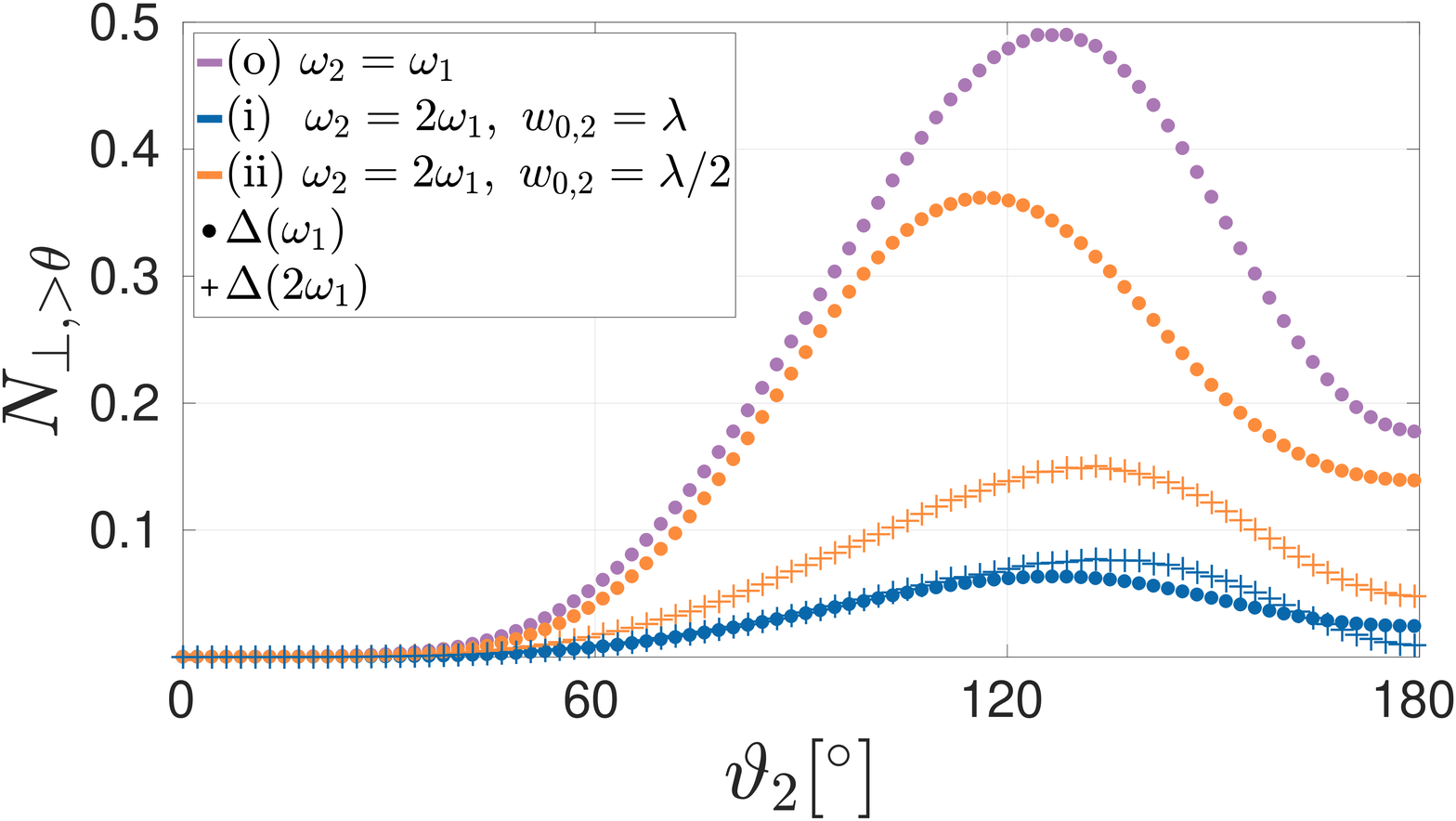}    
\caption{Partitioning of the perpendicularly polarized signal photons
  into the frequency regimes $\Delta(\omega_1)$ and $\Delta(2\omega_1)$
  for the various scenarios (o)-(ii). Both high-intensity laser pulses are polarized perpendicularly to the collision plane. 
  The segment with center
  frequency ${\rm k} =\omega_1$ ($2\omega_1$) is depicted by $\bullet$
  ($+$) symbols.  Naturally, no ${\rm k}\approx2\omega_1$ signal is induced in
  the collision of two fundamental frequency beams. Top: Total
  number of perpendicularly polarized signal photons $N_\perp$. Bottom:
  Integrated number of perpendicularly polarized signal photons
  emitted outside the forward cones of the colliding Gaussian laser
  beams $N_{\perp,>\theta}$.}
\label{fig:Nperp_char}
\end{figure}

Analogously, Fig.~\ref{fig:Nperp_char} shows results for the number of signal photons polarized perpendicularly to the high-intensity laser beams $N_\perp$ and $N_{\perp,>\theta}$.
Besides, in Tab.~\ref{tab:2} we exemplarily stick to a collision angle of $\theta_2=135^\circ$ and provide explicit numerical values for the numbers of signal photons with energies in the ranges $\Delta(\omega_1)$ and $\Delta(2\omega_1)$.
For a given energy regime $\Delta$, the values for $N$ and $N_\perp$ and analogously $n_{>\theta}=\frac{N_{>\theta}}{N}$ and $n_{\perp,>\theta}=\frac{N_{\perp,>\theta}}{N_\perp}$ exhibit similar trends.

Let us first detail on the behavior of $N$ and $N_\perp$.
In the energy regime $\Delta(\omega)$, the largest numbers for $N$ and $N_\perp$ are obtained for scenario (o), followed by (ii) and finally (i).
This is completely consistent with our expectations as the maximum number of frequency-$\omega_1$ signal photons is to be expected for the collision of two fundamental-frequency beams.
As one can see in Fig.~\ref{fig:EmChar1}, these essentially elastically scattered signal photons are predominantly emitted in the forward directions of the high-intensity laser beams.
The finding that the attainable signal photon numbers in scenario (ii) are larger than for scenario (i) hints at the fact that the peak field strength is most decisive for the effect.
Recall, that for (ii) the frequency-doubled laser beam is focused down to the diffraction limit with $f^\#=1$, guaranteeing a maximum peak field, while in (i) it is only focused with $f^\#=2$; cf. Figs.~\ref{fig:beams(i)} and \ref{fig:beams(ii)}.
In the energy regime $\Delta(2\omega_1)$, we find similar trends for the behavior of $N$ and $N_\perp$. Generically, no frequency-$2\omega$ signal is generated in the collision of (o) two fundamental-frequency laser beams; see \Eqref{eq:feqcombs}.

Secondly, we comment on the trends observed for the relative fractions of signal photons $n_{>\theta}$ and $n_{\perp,>\theta}$ scattered outside the beam divergences in forward direction.
Again we first discuss the results obtained for the energy regime $\Delta(\omega_1)$.
This signal is mainly induced in the propagation direction of the high-intensity laser with fundamental frequency, which implies that effectively only the divergence of the fundamental-frequency beam matters.
While the values of $n_{>\theta}$ and $n_{\perp,>\theta}$ are similar for the cases (o) and (i), the result for case (ii) is significantly different.
For the cases (o) and (i), the fundamental frequency beam collides with a beam of similar transverse focus profile of width $w_{0,1}=w_{0,2}=\lambda$.
As the signal photons are predominantly induced in the focus, the similar values obtained for $n_{>\theta}$ and $n_{\perp,>\theta}$ are not surprising.\footnote{Note, that this argument is not invalidated by the fact that in (o) we consider two frequency-$\omega_1$ beams, while there is only a single frequency-$\omega_1$ beam in (i).
The reason for this is the fact that the ratios $n$ are insensitive to the absolute numbers.}

\begin{figure}[t]
\includegraphics[width=\figlenFull]{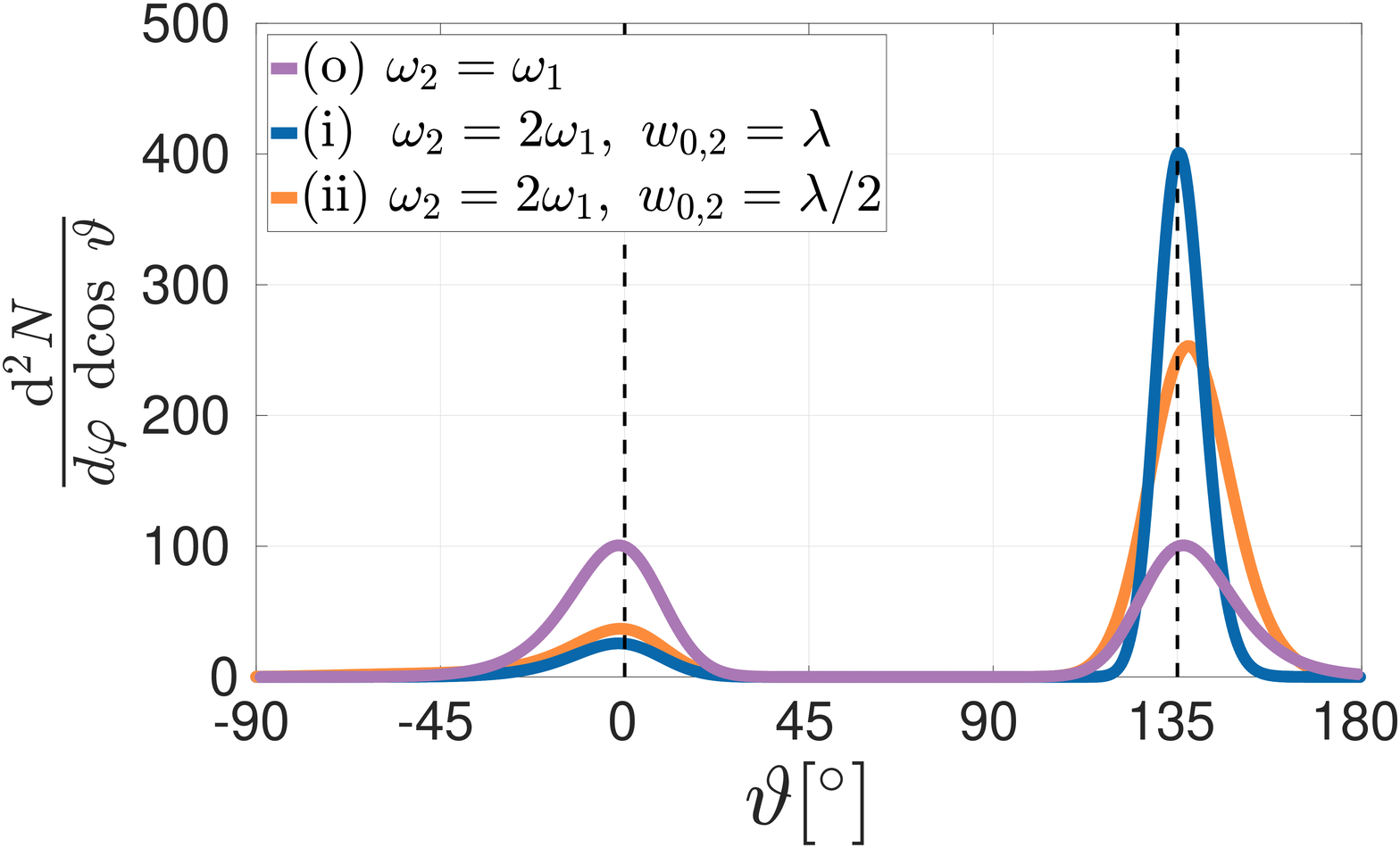}    
\caption{Differential number of signal photons $\frac{{\rm d}^2N}{{\rm
      d}\varphi\,{\rm d}\cos\vartheta}$ for $\varphi=0$, i.e., in the
  collision plane, for a collision angle of $\vartheta_2=135^\circ$,
  plotted as a function of the polar angle $\vartheta$ for the
  scenarios (o)-(ii).  The dashed lines at $\vartheta = 0^\circ$
  ($\vartheta = 135^\circ$) indicate the propagation direction of the
  high-intensity laser beam of frequency $\omega_1$ ($\omega_2$).
  The different peak-widths at $\vartheta \approx 135^\circ$ for
  scenarios (i) and (ii) can be traced back to the different focusing
  of the frequency-doubled laser. Generically, a harder focusing
  results in a wider far-field divergence.}
\label{fig:dN}
\end{figure}

Conversely, the smaller beam waist of the frequency-doubled beam in (ii) naturally gives rise to a larger fraction of photons scattered out of the divergence of the fundamental frequency beam as compared to (o) and (i).
Generically, a tighter scattering center results in a wider angle distribution of the scattered light in the far-field; cf. Ref.~\cite{Karbstein:2016lby} for similar observations in a strong-field QED context.

\begin{figure}[t]
\includegraphics[width=\figlenD]{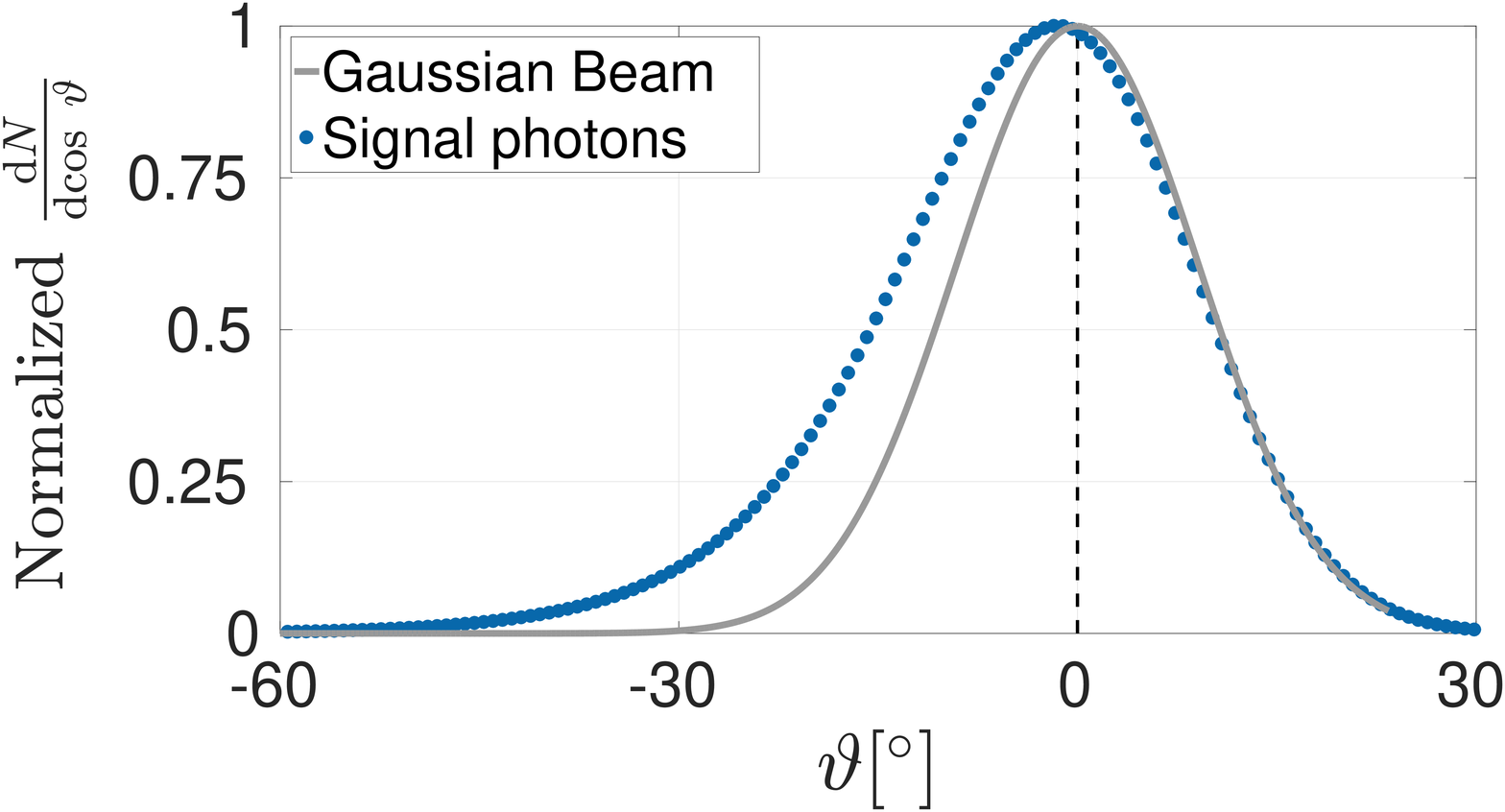}   
\includegraphics[width=\figlenD]{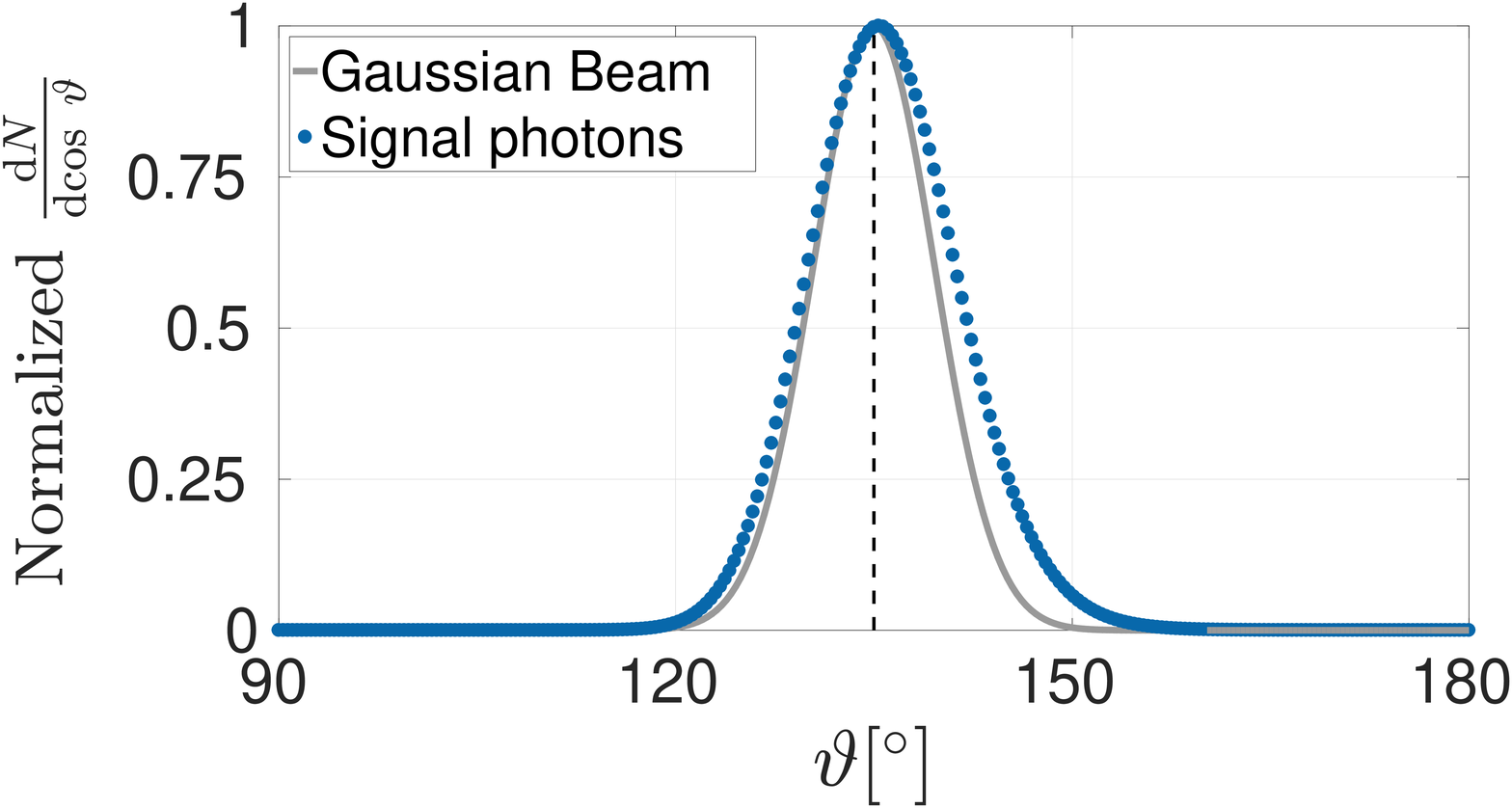}  
\caption{Comparison of the normalized differential number of signal
  photons $\frac{{\rm d}^2N}{{\rm d}\varphi\,{\rm
      d}\cos\vartheta}$ for $\varphi=0$ with the far-field photon
  distributions of the high-intensity laser beams. Here, we
  exemplarily limit ourselves to scenario (i) and a collision angle of
  $\vartheta_2=135^\circ$.  As the frequency-doubled Gaussian beam is only
  focused to $w_{0,2} = \lambda$, its divergence fulfills
  $\theta_2=\theta_1/2$.}
\label{fig:dN_0_135}
\end{figure}

In the energy regime $\Delta(2\omega_1)$, the ordering is reversed, such that the fraction of signal photons scattered out of the divergence of the high-intensity lasers is larger for (i) than for (ii).
This observation can be explained along the same lines as above.
The signal photons with energy in the regime $\Delta(2\omega_1)$ are predominantly emitted in the vicinity of the propagation direction of the frequency-doubled laser beam, implying that the observed trends can be explained by considering the divergence of the $2\omega_1$ beam only.
Now the frequency-doubled beam collides with a fundamental frequency pulse of the (i) same or (ii) wider width; cf. Figs.~\ref{fig:beams(i)} and \ref{fig:beams(ii)}.
Following the reasoning given above, this immediately implies that for (i) more signal photons are expected to be scattered outside the beam divergence of the high-intensity beam than for (ii).

In Fig.~\ref{fig:dN}, we depict the differential number of signal
photons $\frac{{\rm d}^2N}{{\rm d}\varphi\,{\rm
    d}\cos\vartheta}$ at $\varphi=0$ for all three scenarios (o)-(ii)
For the symmetric configuration with two fundamental-frequency beams
(o) both peaks are of the same height, and exhibit a mirror symmetry
with respect to the middle axis between the two beams at
$\vartheta=135^\circ/2=67.5^\circ$; see Fig.~\ref{fig:EmChar1}.  In
the scenarios (i) and (ii) the differential photon number are largest
in the directions of the frequency-doubled beam.

For (o) and (ii) both high-intensity laser beams exhibit the same
divergence $\theta_1=\theta_2$. Conversely, for (i) the divergence of the
frequency-doubled beam is $\theta_2=\theta_1/2$, which explains why
for (i) also the signal photons are scattered into a narrower
far-field angle.

To allow for a comparison of the angular spread of the photons
constituting the high-intensity laser beam and the signal photons, we
plot the corresponding differential photon numbers in the far-field as
a function of the polar angle $\vartheta$ in Fig.~\ref{fig:dN_0_135}.
The photon distributions of the high-intensity laser beams in the
far-field scale as $\frac{{\rm d}^2N}{{\rm d}\varphi\,{\rm
    d}\cos\vartheta} \sim {\rm e}^{- 2 \vartheta^2/ \theta_1^2 }$ for
the beam propagating along the $\rm z$ axis and as $\frac{{\rm
    d}^2N}{{\rm d}\varphi\,{\rm d}\cos\vartheta} \sim {\rm e}^{-2
  \left( \vartheta-\vartheta_2 \right)^2/(\theta_1/f^\#
  )^2 }$ for the other beam; where $f^\#=1$ for both (o) and
(ii), and $f^\#=2$ for (i).  Obviously, the signal photons are
scattered asymmetrically.  The different decay of the signal photons
and the photons constituting the high-intensity laser fields leads to
an improved signal to background ratio.

\section{Conclusions and Outlook}\label{sec:concl}

In this article, we have provided further evidence that all-optical
signatures of quantum vacuum nonlinearity can be analyzed efficiently
in terms of vacuum emission processes. The essence of this concept is
that all macroscopically sourced fields are treated as classical,
whereas the fields induced by quantum nonlinearities receive a quantum
description in terms of signal photons. This concept matches ideally
with the physical situation and thus provides direct access to
physical observables.

In the present example of colliding laser pulses, this approach
facilitates to directly determine the directional emission
characteristics and polarization properties of the signal photons
encoding the signature of quantum vacuum nonlinearities.  Our main
goal was to demonstrate that, assisted by a dedicated numerical
algorithm, the vacuum emission approach is particularly suited to
tackle signatures of strong-field QED in experimentally realistic
electromagnetic field configurations generated by state-of-the-art
high-intensity laser systems.  To this end, we focused on a
comparatively straightforward scenario, based upon the collision of
two optical high-intensity laser pulses, which we model as pulsed
Gaussian beams.  Resorting to a locally constant field approximation
of the Heisenberg-Euler effective action, our numerical algorithm
allows for a numerically efficient and reliable study of the
attainable numbers of signal photons for arbitrary collision angles
and polarization alignments.
Our formalism can be readily extended to the collision of more laser beams, such as the study of photon-merging \cite{Gies:2016czm}, or equivalently four-wave mixing processes \cite{Lundstrom:2005za,Lundin:2006wu} induced by QED vacuum nonlinearities in the collision of three focused high-intensity laser beams.

\section*{Acknowledgments}

We are grateful to Nico Seegert for many helpful discussions and
support during the development phase of the numerical algorithm.  The
work of C.K.~is funded by the Helmholtz Association through the
Helmholtz Postdoc Programme (PD-316). We acknowledge support by the
BMBF under grant No. 05P15SJFAA (FAIR-APPA-SPARC).

Computations were performed on the ``Supermicro Server 1028TR-TF'' in Jena, which
was funded by the Helmholtz Postdoc Programme (PD-316).

\appendix
\section{Convergence tests}
\label{App}

As discussed in the main text, semi-analytical and numerical results
fit almost perfectly for a suitable choice of numerical discretization
parameters. In the following, we detail this choice of numerical
parameters by studying the convergence of the numerical algorithm in
comparison to the semi-analytical results for the toy-model benchmark
test.  Such an analysis is useful, because it (i) helps to improve the
stability of the numerical results and (ii) yields systematic checks
enabling to run simulations in regions of the parameter space, where
no analytical reference values are available. Eventually, it also
helps to minimize the program's runtime as well as its memory
requirements.

In this work, we have in total $10$ independent parameters controlling
the numerical calculation. These are $N_{\rm x}$, $N_{\rm y}$, $N_{\rm
  z}$ specifying the lattice in the Cartesian grid for
spatial/momentum coordinates, $N_{\varphi}$, $N_{\vartheta}$, $N_{\rm
  k}$ yielding the number of grid points in spherical momentum
coordinates and $L_{\rm x}$, $L_{\rm y}$, $L_{\rm z}$, $L_{\rm k}$
defining the physical interval of length $2L_{\rm x,y,z,k}$ (sampling
regions) of the corresponding variables centered around the region of
interest. For illustration, we focus here on lower dimensional
subsets. Similar convergence checks can be performed for each of these
parameters.

\begin{figure}[t]
\includegraphics[width=\figlenFull]{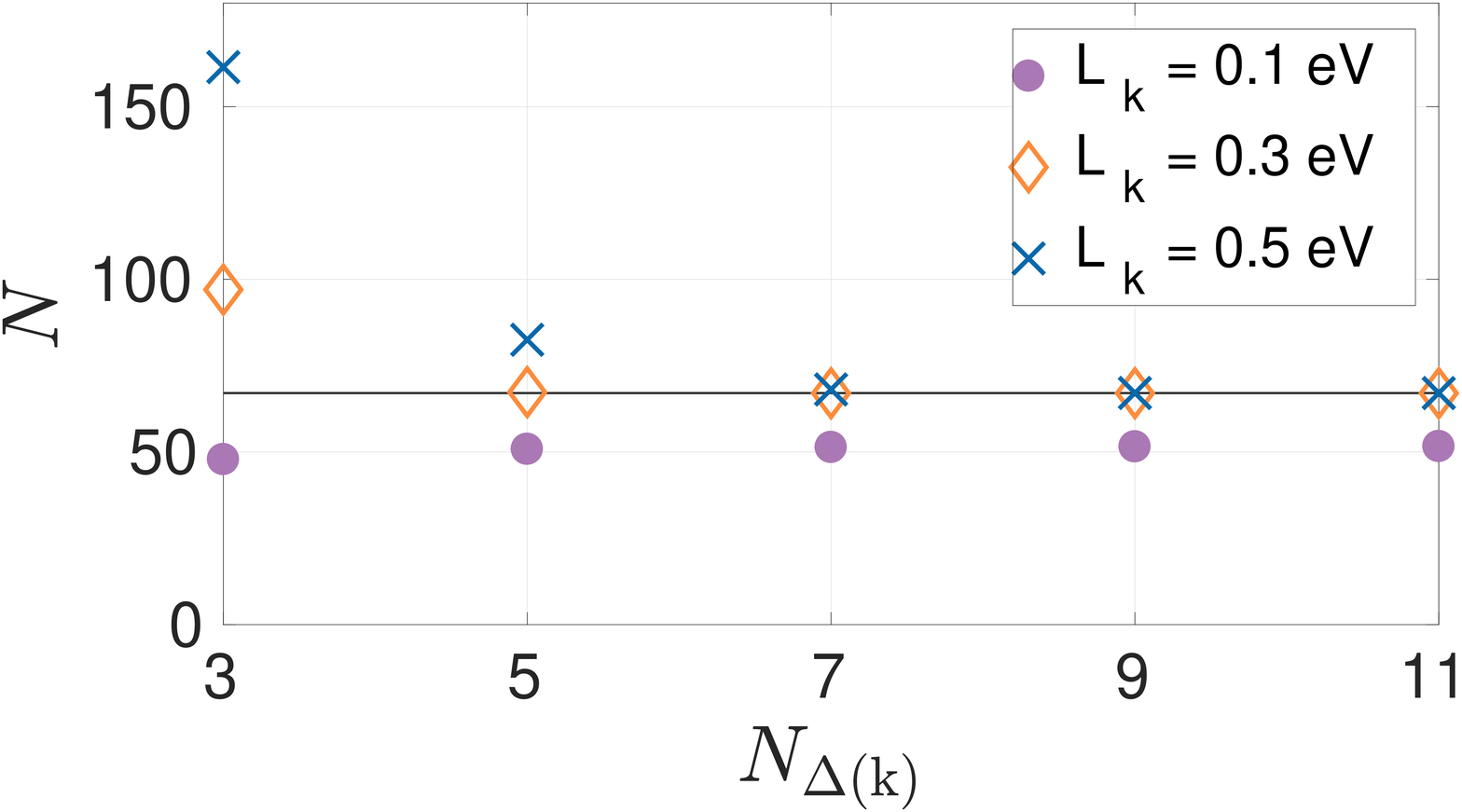}      
\caption{Convergence of the total number of signal photons $N$ as a
  function of the number of radial momentum grid points in $N_{
    \Delta({\rm k})}$ for various sizes of the sampling region.  The
  sampling interval in ${\rm k}$ has a total length of $2L_{\rm k}$
  and is centered around $\omega \approx 1.54 {\rm eV}$.  Two
  identical laser pulses ($\lambda=800{\rm nm}$, $W=25{\rm J}$,
  $\tau=25{\rm fs}$) are focused to $w_{0,1}=w_{0,2}=\lambda$. They
  are assumed to be polarized perpendicularly to the collision plane
  and collide under an angle of $\vartheta_2=135^\circ$.  The
  benchmark toy model is used here to allow for a comparison with the
  semi-analytical result (black line).}
\label{fig:Conv_k}
\end{figure}

In the following, we discuss the numerical convergence of our
calculations in the context of two parameters, the radial momentum of
the signal photons ${\rm k}$ and the longitudinal resolution of the
pump fields along the ${\rm z}$ axis. For this, we first plot the
total number of signal photons $N$ as a function of the number of grid points
$N_{\Delta({\rm k} )}$ for various choices of the momentum grid
length $L_{\rm k}$ in Fig. \ref{fig:Conv_k}. By comparison with the
semi-analytical results we observe, that accuracy of the result
increases with the momentum-space resolution as expected. It is also
remarkable, that a few grid points in the total momentum ${\rm k}$,
$\mathcal O (10)$, are sufficient in order to approximate the
analytical solution reasonably well. The crucial ingredient is, of
course, an appropriate choice for the resolved momentum interval:
while the center of the $L_{\rm k}$ region can be adapted to the
requirements imposed by energy conservation, cf. \Eqref{eq:feqcombs},
being $k\simeq \omega$ in the present example, the size of $L_{\rm k}$
has to cover the bandwidth of the outgoing pulse. In the present case,
a region with $L_{\rm k}\geq 0.3$eV is required, corresponding to
$\geq 20\%$ of the central pulse energy. For instance, a region
limited to $L_{\rm k}=0.1 {\rm eV}$ is not sufficient to provide a
precise estimate of the signal photon number, see
Fig. \ref{fig:Conv_k}.

Secondly, we investigate the spatial resolution needed in order to
satisfactorily resolve the applied laser pulses. In this case, the
parameters $L_{\rm z}$ and $N_{\rm z}$ have to meet two different
requirements: on the one-hand side, $L_{\rm z}$ has to be chosen large
enough to cover the region of interest given by the focal and
collision region of the two pulses, while $N_{\rm z}$ has to be
sufficiently large to precisely sample the details of the pulse shape;
On the other hand, the nature of the Fourier transform implies that
$\pi/(2L_{\rm z})$ defines an infrared cutoff and $\pi N_{\rm
  z}/(2L_{\rm z})$ an ultraviolet cutoff for the ${\rm z}$ component
of the momentum of the outgoing signal photon. Hence, both have to be
chosen sufficiently large also to resolve the sampling region $2L_{\rm
  k}$ centered around the peak momentum ${\rm k}$ of the signal photon
appropriately. As a rule of thumb, an increase of the sampling region
should go along with an increase of the number of grid points in order
to keep the momentum space ultraviolet resolution (at least) constant.

In the present case, the procedure for choosing the discretization
parameters is the following: The parameter $L_{\rm z}$ should be
chosen large enough in order to resolve the focal region of the pump
fields, i.e. at least one oscillation of the pump fields in the
present case. Signal energy conservation suggests the signal photons
to be located at around ${\rm k} \approx \omega$, the values for
$L_{\rm z}$ and $N_{\rm z}$ should take on values such that the
momentum region around $\omega$ is with sufficient resolution within
the infrared and ultraviolet cutoffs induced by the Fourier
transformation.  For definiteness, we have fixed the longitudinal
sampling region to ${\rm z} \in 2^q [-0.95 \lambda, 0.95 \lambda]$ and
study the convergence of the result for increasing $q$ and $N_{\rm z}$.

\begin{figure}[t]
\includegraphics[width=\figlenFull]{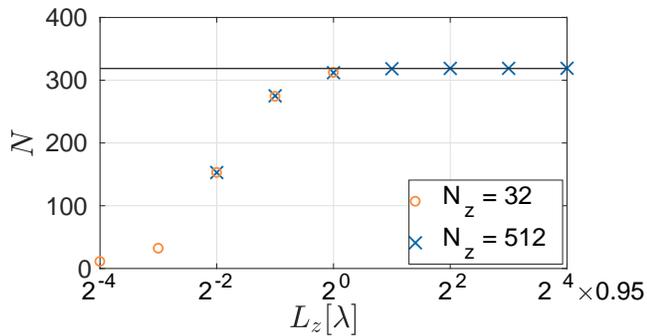}      
\caption{Convergence of the total number of signal photons $N$ as a
  function of the interval length $[-L_{\rm z}, L_{\rm z} ]$ for two
  sets of grid points in the ${\rm z}$ direction.  Two identical laser
  pulses ($\lambda=800{\rm nm}$, $W=25{\rm J}$, $\tau=25{\rm fs}$) are
  focused to $w_{0,1}=w_{0,2}=\lambda$ and collide in a
  counter-propagation geometry.  Both pulses are polarized
  perpendicularly to the collision plane.  The benchmark toy model is
  used here to allow for a comparison with the semi-analytical result
  (black line).}
\label{fig:Conv_z}
\end{figure}
  
The results for the signal photon number as a function of the size of
the sampling region for two different grid resolutions are shown in
Fig. \ref{fig:Conv_z} and Tab. \ref{tab:App1}. As expected, the
spatial sampling region has to be large enough to cover the focal
region of the size of a wavelength $\lambda$ in order to approach the
correct result. We observe that even a rather small number of $32$
grid points can give an acceptable result with an error on the percent
level, if the size of the sampling region is chosen appropriately as
to cover the relevant momentum region of the signal photon upon
Fourier transformation. For a reliable result with an error well below
1\%, larger numbers of grid points and a sufficiently large sampling
region are required -- of course, at the expense of computing time.

\setlength{\extrarowheight}{3pt}
\setlength\tabcolsep{12pt}
\begin{table}[t]
  \caption{Benchmark calculations for the total number of signal photons attainable in the toy-model scenario. Two identical laser pulses ($\lambda=800 {\rm nm}$, $W=25{\rm J}$, $\tau=25{\rm fs}$), focused to $w_{0,1}=w_{0,2}=\lambda$ and polarized perpendicularly to the collision plane, collide under an angle of $\vartheta_2 = 180^\circ$. 
    The overall runtime and the corresponding mean relative error $\text{MRE}_{N}$ with respect to the  semi-analytical result ($N = 330.189$) are listed as  functions of the grid size $L_{\rm z}$ and the number of grid points $N_{\rm z}$.}
\begin{center}
\begin{tabular}{lcccc}
 \toprule
  & & \multicolumn{3}{c}{Mean rel. error $\text{MRE}_{N}[\%]$} \\  
  \cmidrule(lr){3-5}
  & & \multicolumn{3}{c}{Grid size $L_{\rm z}$ [$0.95 \ {\rm \lambda}$]} \\
  \cmidrule(lr){3-5}
  $N_{\rm z}$ & Runtime [s] & $2^{-1}$ & $2^0$ & $2^1$ \\ 
\hline
 32 & 1120 & 15.25 & 2.17 & - \\ 
 128 & 3105 & 14.87 & 2.20 & 0.22 \\ 
 512 & 9400 & 14.86 & 2.20 & 0.22 \\ 
\end{tabular}
\end{center}
\label{tab:App1}
\end{table}
\setlength{\extrarowheight}{0pt}
\setlength\tabcolsep{0pt}

\FloatBarrier

\end{document}